\documentstyle[12pt]{article}

\def\head{\vspace*{3cm}}
\begin{document}

\newcommand{\ba}{\begin{array}}
\newcommand{\ea}{\end{array}}
\newcommand{\bd}{\begin{displaymath}}
\newcommand{\ed}{\end{displaymath}}
\newcommand{\be}{\begin{equation}}
\newcommand{\ee}{\end{equation}}
\newcommand{\bea}{\begin{eqnarray}}
\newcommand{\eea}{\end{eqnarray}}
\def\eq{}
\def\eqs{}
\def\Eq{Equation~}
\def\Eqs{Equations~}
\def\L{{\cal L}}
\def\eff{{\rm eff}}
\def\red{{\rm red}}
\def\Lo{{\cal{L}_{\rm o}}}
\def\Ro{{R_{\rm o}}}
\def\Leff{{\cal{L}_{\rm eff}}}
\def\Lred{{\cal{L}_{\rm red}}}
\def\LQCD{{\cal{L}_{\rm QCD}}}
\def\EOM{{\rm EOM}}
\def\OEOM{O_{\rm EOM}}
\def\OBRS{O_{\rm BRS}}
\def\O{{\cal{O}}}
\def\OP{\cal{O}_{\rm P}}
\def\OB{\cal{O}_{\rm B}}
\def\OE{\cal{O}_{\rm E}}
\def\Y{Q}
\def\geff{g_{\rm eff}}
\def\dEOM{\d_{\rm E}}
\def\dy{dy}		 % integration
\def\dd{\delta^{(d)}}	 % Dirac-delta
\def\i{i}                % math-i
\def\ln{\mbox{$\ell n$}} % logarithm
\def\to{\rightarrow}     % decay into
\def\too{\longrightarrow}% limes
\def\exp{{\rm exp}}	 % exp
\def\log{\ell og}	 % log
\def\ln{\ell n}	 	 % ln
% Bra-Kets:
\def\bra{\langle}
\def\ket{\rangle}
% Spinors:
\def\Psib{b}
\def\Psis{\bar{s}}
\def\Psibar{\overline{\Psi}}
% Slashes:
\def\ds{\mbox{$\not\!\partial$}}
\def\Ds{\mbox{$\not\!\! D$}}
\def\As{\mbox{$\not\!\! A$}}
\def\ds{\mbox{$\not\!\partial$}}
\def\ks{\mbox{$\not\!k$}}
\def\ps{\mbox{$\not\!p$}}
\def\qs{\mbox{$\not\!q$}}
% Bars on quarks:
\def\ub{\bar{u}}
\def\db{\bar{d}}
\def\cb{\bar{c}}
\def\sb{\bar{s}}
\def\qb{\bar{q}}
% Greek letters:
\def\mn{{\mu\nu}}
\def\a{\alpha}
\def\b{\beta}
\def\g{\gamma}
\def\d{\delta}
\def\e{\epsilon}
\def\ve{\varepsilon}
\def\et{\eta}
\def\k{\kappa}
\def\l{\lambda}
\def\m{\mu}
\def\n{\nu}
\def\o{\omega}
\def\r{\rho}
\def\s{\sigma}
\def\D{\Delta}
\def\G{\Gamma}
\def\S{\Sigma}
% Fractions:
\def\2{{1\over 2}}
\def\3{{1\over 3}}

\begin{flushleft}
DESY 93-083\\
hep-ph/9307274
\end{flushleft}
\head
\begin{center}
{\large \bf{Equations of Motion for Effective Lagrangians}}\\
{\large \bf{and Penguins in Rare B-Decays}}
\vspace*{2cm}\\
{\large H. Simma}\\
Deutsches Elektronen-Synchrotron DESY, Hamburg
\end{center}
\vspace*{2cm}
\begin{abstract}

\noindent
We study the application of the classical equations of motion (EOM)
within the framework of an effective low-energy Lagrangian
treated at the loop level. Gauge-fixing and ghost terms, which
enter naturally in the EOM, are found to lead to no physical effects
--- neither through operator mixing nor in matrix elements.
Beyond first order in the effective interactions, contact terms
have to be included when reducing the effective Lagrangian and we present
an explicit procedure to construct them.
Applied to (hadronic) rare B-decays, the EOM drastically simplify the
effective Lagrangian and its matching to the underlying theory, and
certain cancellations of large (logarithmic) contributions become more
transparent. Finally, we discuss details of the `matching' of the
effective Lagrangian, which may be helpful in incorporating short
distance QCD corrections in further phenomenological studies.
\end{abstract}
\clearpage

\section{Introduction}
Processes in which the external momenta and masses are much smaller than
the mass scale of the particles that mediate the interactions, are
most conveniently described in the framework of an effective field theory
\cite{EFT}. There, the heavy particles are integrated out and their
relevant effects are summarized in a concise way by local operators in an
effective Lagrangian. Moreover, short-distance (QCD) corrections can
be included in a systematic way by means of the renormalization group.
Typical phenomenological applications of effective theories are
quantitative predictions for low-energy processes on the basis of
a given theory (e.g. the standard model) or when
experimental data is analyzed in order to obtain bounds or hints
for yet unknown `new physics' (e.g. through anomalous gauge-boson
couplings).

The complete set of linearly independent effective operators
for certain physical processes usually contains many operators which
are formally related by the classical equations of motion (EOM).
It is desirable to exploit these relations among the various operators
to simplify the expression for a {\it given} effective
Lagrangian\footnote{
   i.e. a linear combination of effective operators
   with {\it fixed} coefficients.
   Note the difference to the question of simply reducing the set of
   operators to describe general on-shell amplitudes. For instance, it
   has been stressed \cite{Arzt} that operators containing covariant
   derivatives, $\Ds\Psi$, $D_\m G^\mn$, etc., can be dropped from the
   complete set of operators.
   However, the explicit form of the relation between
   the operators is not specified; rather their coefficients are to be
   determined by a complicated matching procedure where many additional
   1PR graphs must be evaluated.}
and at any stage of the calculation (e.g. after integrating out heavy
particles and before performing the renormalization group evolution).
To this end, we investigate in this paper the explicit form of these
relations and their preservation by renormalization effects.

Of course, at tree (i.e. classical) level, on-shell matrix elements
of an effective operator may be simplified by using the classical EOM,
which include the interactions from the usual, e.g. QCD$\times$QED,
Lagrangian (to account for the fact that internal lines entering in
the effective operator may actually be off shell).
When working at higher order in the {\it effective} interactions,
a more careful formulation of the EOM, which will also depend
on the effective interactions themselves, is necessary.
(Typically, an effective Lagrangian is applied only at first
order, but one may think of processes, like $K^o-\bar{K}^o$ or
$B^o-\bar{B}^o$ mixing, which arise only at second order in the
effective weak interactions.)

At the loop level also renormalization effects have to be taken into
account, and they are, in fact, crucial for assessing the applicability
of the EOM {\it before} the renormalization group (RG) evolution.
The renormalization of composite operators is well-known and has been
studied in great detail \cite{Zim,EOM,Jog,Shore}; however, in
phenomenological work the issue of the EOM is usually either not
discussed in a coherent way or the general results are not exploited.

A complication, which is sometimes overlooked, arises from the fact
that the appropriate Lagrangian for the quantized theory contains
gauge-fixing terms and Faddeev-Popov ghosts. A priori, these terms are
to be taken into account in the EOM and they can, in principle, contribute
to physical processes through mixing with physical operators and/or through
non-vanishing matrix elements at the loop level. The crucial point is
that the additional gauge-fixing and ghost terms enter the EOM in a
combination, which is the BRS variation of some other operator; then,
using standard arguments \cite{Jog,Collins}, we find that these
additional terms do not affect the matrix elements for physical processes.
Throughout this paper we shall use the convention that the `classical
EOM' include possible gauge-fixing and ghost contributions --- in
contrast to the `naive classical EOM' without these terms.

While our discussion of the EOM can be useful in various other
phenomenological studies (for instance, of anomalous gauge-boson
couplings), we focus our attention in the second part of this paper
on hadronic rare B-decays . In particular, we consider here
flavour changing neutral $b\to s$ currents which arise from loop
diagrams, so called `penguins', with internal $u$, $c$ or top quark.
The corresponding tree diagrams --- if at all present --- are highly
CKM suppressed.

Penguin diagrams do not only induce interesting exclusive decays, like
$B\to K^\ast \g$ \cite{CLEO}, but also contribute considerably
to the inclusive hadronic branching ratio for $b\to s + no\ charm$
(via $b\to sq\qb$, $b\to sgg$, etc. \cite{Hou}), and they can generate
CP violating asymmetries in charged B-decays \cite{BSS}.
In various of these processes, one finds a remarkable cancellation \cite{gg0}
of large --- since only weakly GIM suppressed --- contributions with
logarithmic dependence on the internal mass.
We demonstrate how this cancellation, which in fact
has been missed in the pioneering works \cite{Hou,GH},
is made transparent by the EOM.

The EOM are particularly interesting for
calculations of short-distance QCD effects (which lead to sizable
corrections in the branching ratios \cite{BBM} and asymmetries
\cite{ex0,Fleischer}).  In the abundant literature on such
calculations for K- and B-meson decays, essentially two approaches are
to be found: Many authors perform the RG evolution with an effective
Lagrangian not reduced by the EOM (see for instance
refs.~\cite{AMP,GODS,Cella}). Of course, this is a rather tedious
procedure because the corresponding anomalous dimension matrix is
large. In some cases it has been observed that the final result for
the process considered is independent of whether the EOM are applied
before or after performing the RG evolution.  While in
ref.~\cite{GODS} this fact has been interpreted as a nontrivial result
of the calculation, the authors of ref.~\cite{AMP} suspected that this
is not an accidental fact, but did not give a general argument.

Other authors, for instance, Grinstein et al. \cite{GSW} and Buras et al.
\cite{Buras}, employ the {\it naive} classical EOM to simplify the
operator basis right from the beginning.  The applicability of these
EOM is either not discussed explicitly or assumed on the basis of a
formal proof by Politzer \cite{Pol} (which by a close inspection actually
leads to the classical EOM including the gauge-fixing and ghost terms).
We shall clarify this point by demonstrating the appearance of
the gauge-fixing and ghost contributions in terms of diagrams and
by investigating their renormalization.

Finally, we would like to discuss some details concerning the `matching'
of the full theory with the effective Lagrangian (reduced by the EOM),
and the recovery of the momentum dependent and absorptive parts of the
amplitudes in the effective theory at leading and next-to-leading log
precision. These issues are often somewhat obscured in the literature
by other technical details related to the actual calculation of the
RG evolution.

The remainder of this paper is organized as follows: In section~2
we review the proof of the EOM and verify the absence of physical
effects from the gauge-fixing and ghost terms. We then discuss the
appropriate form of the EOM when working at higher order in the effective
Lagrangian. In section~3 we exploit the EOM to reduce the effective
penguin operators in $b\to s$ transitions, thereby taking inventory of
the coefficients with logarithmic mass dependence, and we describe the
`matching' of the effective Lagrangian. The discussion in section~4
contains some clarifying remarks, and additional details, which may
be helpful for further applications, are collected in the appendices.

\section{Equations of motion}
\subsection{General framework}
The Lagrangian of an effective theory \cite{EFT} consists of two parts:
\bd \cal{L} = \Lo + \Leff \ .\ed
$\Lo$ is renormalizable by power
counting and describes those interactions of the particles which, in
principle, are to be taken into account at arbitrary order in
perturbation theory. Of course, for a consistent quantization $\Lo$ has
to {\it include} terms for the gauge-fixing and Faddeev-Popov ghosts.
The effective Lagrangian, $\Leff$, is a linear
combination of higher dimensional local operators representing the
effective interactions. They have dimensionful couplings and are to be
treated only up to a certain order, usually the first one.  The
effective interactions may be thought as being induced by heavy
particles of an underlying more fundamental theory.  A typical example
is $\Leff$ arising from effective weak interactions (induced by virtual W's
and top quarks) with $\Lo = \LQCD$ being the usual QCD Lagrangian
(neglecting QED interactions for simplicity).

In order to regularize UV divergences in calculations beyond the tree
level we will always assume dimensional regularization.
The composite operators, $O_i(y)$, in $\Leff$
are defined \cite{Zim,Collins} by
their Green's functions $\bra T O_i(y) X(x_1,\ldots)\ket$, where $T$
indicates time-ordering and the fields are interacting with respect to
$\Lo$.  Physical matrix elements or operator identities are obtained
by LSZ reduction, and the Green's functions themselves are defined
perturbatively by (regularized) Feynman graphs. In the following, $X$ will
always denote a product of {\it elementary} fields at different points
\bd X(x_1,\ldots)\equiv \Phi(x_1)\ldots A^{a_1}_{\m_1}(x_{k+1})
\ldots \bar{\Psi}(x_{m}) \ldots\Psi(x_{n}) \ldots \ . \ed

For fixed values of the regularization parameter $\epsilon$ and of the
renormalization scale $\mu$, the renormalized effective theory
is specified by three elements: (i) The
Lagrangian, $\L(g,\cdots;\Phi)$, as a function of the fields and parameters
(like couplings and masses), (ii) the values ${g^R,\ldots}$ for
these parameters, and (iii) the counter term Lagrangian, $\L^{\rm ct}$ ---
or, equivalently, a renormalization prescription to determine $\L^{\rm ct}$
from $\L$. We call $\L^{\rm basic}(\Phi) \equiv \L(\m^\e g^R,\ldots;\Phi)$
the `basic' part of the `renormalized' Lagrangian
$\L^R(\Phi) \equiv \L^{\rm basic}(\Phi)
+ \L^{\rm ct}(\m^\e g^R,\ldots,\epsilon;\Phi)$, where the usual powers of
$\mu^\e$ keep the dimensions of the $g^R$ independent of $\epsilon$.
In the following, we will always assume minimal (or $\overline{ms}$)
subtraction for the renormalization prescription.

Renormalized Green's functions are calculated by using $\L^R$ and remain
finite for $\epsilon \longrightarrow 0$ due to suitable counter
terms in $\L^{\rm ct}$. While all counter terms needed for Green's
functions of elementary fields are proportional to terms in $\Lo$
(multiplicative renormalization of $\Lo$), the renormalization of
insertions of composite operators requires additional subtractions
proportional to other composite operators. Since $\Leff$
is to be kept only up to a finite order, only a finite number of such
counterterms is needed; in this sense $\Lo+\Leff$ is renormalizable.
To render the effective operators multiplicatively renormalizable, one
includes a priori all interactions in $\Leff$ that may appear as
counterterms.

The couplings of $\Leff$, i.e. the coefficient functions of the
operators, are determined by `matching' on-shell amplitudes either
with a more fundamental theory or with experimental data. By this
procedure no distinction is possible between effective theories which
yield the same on-shell amplitudes. Henceforth, different Lagrangians
yielding the same on-shell amplitudes will be called `on-shell equivalent'.
An `on-shell effective theory' \cite{EFT} can be viewed as a class of
on-shell equivalent effective theories.

If two (effective) theories, renormalized at any scale $\mu$,
are on-shell equivalent, then the corresponding `bare' theories\footnote{
   with bare parameters, ${g_0(\e),\ldots}$, which are divergent functions
   of $\epsilon$, such that the regularized bare Green's functions,
   calculated from the bare Lagrangian
   $\L^{\rm bare}(\Phi)\equiv \L(g_0(\e),\ldots;\Phi)$,  differ from the
   renormalized ones only by field-renormalization factors.}
are also on-shell equivalent, and vice versa. Since the renormalization
group  describes the relation between the parameters of the various
renormalized theories corresponding to the same bare theory but different
values of $\mu$, the RG evolution preserves the on-shell
equivalence of effective Lagrangians.  Consequently, one can simplify
the expression for $\Leff$ already {\it before} the RG evolution by
replacing $\Leff$ by an on-shell equivalent `reduced' effective Lagrangian.

To study how the {\it equations of motion} can be exploited in order to
obtain on-shell equivalent effective Lagrangians, we consider a generic
effective operator $\OEOM$, which contains the fields in a combination
that vanishes by the classical EOM derived from $\Lo$:
\be\OEOM(z) = Q(z) \cdot \left(\frac{\d\Lo}{\d\Phi(z)}
- \partial_\mu \frac{\d\Lo}{\d\partial_\mu \Phi(z)}\right)\ ,\label{Oeom}\ee
where $\Phi$ represents one of the fields from $\Lo$, and $Q$ is a
monomial in any of the fields and their derivatives (at the same point).
Of course, we exclude the case $Q=\Phi$, since then $\OEOM$ would
correspond to an operator in $\Lo$ and, in particular, would contain the
kinetic energy term. For Green's functions with an insertion of $\OEOM$
one finds \cite{Zim,EOM} (after integration over $z$ to avoid derivatives
of Dirac delta functions)
\be\ba{l} \displaystyle
\int \bra T \OEOM(z) O_1(y_1) \ldots O_n(y_n) X(x_1,\ldots)\ket dz=
\nonumber \\ \hspace*{3cm}
{\displaystyle \i \int \bra T \: \sum_\Lambda \Lambda[Q(z)] \:
\frac{\d}{\d \Lambda[\Phi(z)]} }
\Bigl(O_1(y_1) \ldots O_n(y_n) X(x_1,\ldots) \Bigr) \ket dz\ ,
\label{eom} \ea\ee
where $\Lambda[f] \equiv f, \partial_\m f, \partial_\m\partial_\n f,\ldots$,
denotes the various derivatives (including the fields themselves)
that may enter in the composite operators $O_i$.

For bare Green's functions eq.~(\ref{eom}) is readily derived
by inspection of the contributing Feynman diagrams [see subsection 2.2
for an illustration]. The r.h.s. of \eq(\ref{eom}) consists of
{\it contact terms} which contribute only when $z$ coincides with one
of the $x_1,\ldots$ or $y_1,\ldots,y_n$. They originate from diagrams where
the inverse propagator in $\OEOM$ cancels either an external line
from one of the elementary fields in $X(x_1,\ldots)$, or an internal
line ending up in a vertex generated by one of the other composite
operators $O_1(y_1) \ldots O_n(y_n)$, respectively.

\Eq(\ref{eom}) is also valid for the renormalized Green's functions;
the derivatives $\d/\d\Phi$ etc. in \eqs(\ref{Oeom}) and (\ref{eom})
are then to be taken with respect to the renormalized fields and $\Lo$
refers only to the basic part of the Lagrangian. The renormalized version
of \eq(\ref{eom}) is shown either in terms of diagrams \cite{EOM},
or using the multiplicative renormalization of the composite operators
by induction in the number of loops and insertions.

\subsection{EOM at first order in $\Leff$}
In many applications of effective theories the interactions from
the effective Lagrangian are to be treated only at {\it first} order.
Therefore, we restrict ourselves with the exception of subsection~2.5
to this case. We write
\be \L_{\rm eff}^R = \geff \sum c_i O_i + {\rm counterterms}\ ,
\label{Leff1} \ee
where $\geff$ collectively denotes the dimensionful couplings (for instance,
$\geff = G_F$ in the case of effective weak interactions). All quantities
to be considered are at most linear in $\geff$ or, equivalently, in the
coefficient functions $c_i$.

In particular, only Green's functions with single insertions of the
composite operators from $\Leff$ are needed, and \eq(\ref{eom})
becomes particularly simple
\be\ba{l}
\int\bra T\OEOM(z)\;\Phi(x_1)\ldots\Phi(x_k)\hat{X}(x_{k+1},\ldots)\ket dz =
\nonumber \\ \hspace{2cm}
{\displaystyle\i\sum_{j=1}^k}\:\bra T \Phi(x_1)\ldots
\Phi(x_{j-1})\;Q(x_j)\;\Phi(x_{j+1})\ldots\Phi(x_k)\hat{X}(x_{k+1},\ldots)
\ket \ , \label{eom0}\ea\ee
where $\Phi$ denotes the field to which the variation in \eq(\ref{Oeom})
refers, and $\hat{X}$ is a product of elementary fields containing no
$\Phi$'s. The contact terms on the r.h.s. do not survive the LSZ reduction
and, therefore, do not contribute to matrix elements of physical processes
(see e.g. Joglekar in ref.~\cite{Jog}, and \cite{Pol} for a discussion of
this issue beyond the parton level). Hence, any multiple of $\OEOM$
can simply be dropped from $\Leff$.

Inspecting the reasoning by which \eq(\ref{eom0}) is
derived for the renormalized Green's functions, one notes that the
counterterms for the renormalization of $\OEOM$ are
themselves proportional to operators which vanish by the classical EOM.
Of course, the counterterms are in general not proportional to $\OEOM$
itself, but in fact related to counterterms for $Q$. For instance, in
$\Phi^3$ theory, $\OEOM = \Phi^3(\partial^2 \Phi + m^2 \Phi + g
\Phi^2/2)$ requires counterterms proportional to
$\Phi^2(\partial^2 \Phi + m^2 \Phi + g \Phi^2/2)$ etc.

The derivation of \eq(\ref{eom}) or (\ref{eom0}) in the path integral
representation by a change of variables
is not very intuitive, and might have even been
misleading in cases when ref.~\cite{Pol} is quoted in order to justify
the naive classical EOM for the gluons (without gauge-fixing and ghost
terms). To illustrate the derivation in terms of the contributing Feynman
diagrams we consider here the EOM for the gauge-boson (gluon) field strength
\be D_\mu G^{\mu\nu}_a \equiv
\left(\partial_\m \d_{ab} - g_s f_{abc} A_\m^c \right) G_b^{\m\n} =
\left( J_{\qb q}  + J_{\rm gf} + J_{\rm FP} \right)^\n_a \ . \label{eom2}
\ee
While in the naive classical EOM only the quark contribution
\be \left(J_{\qb q}\right)_a^\nu =
- g_s \sum_{quarks} \qb \g^\nu \frac{\l^a}{2} q \ , \label{Jqq} \ee
is present, the two other terms on the r.h.s.,
$J_{\rm gf}$ and $J_{\rm FP}$, arise in an unambiguous way
from the gauge-fixing and ghost terms in the Lagrangian.

In a covariant gauge (for expressions in a background gauge see appendix~A)
the gauge-fixing and Faddeev-Popov parts of the Lagrangian are given by
\be
{\cal L}_{\rm gf} = - \frac{1}{2\xi} (\partial A)^2
\ , \ \ {\rm and}\ \ \
{\cal L}_{\rm FP} = - (\partial_\m \bar{\eta}_a)  D^\m_{ab} \eta_b \ , \ee
respectively, and the resulting currents on the r.h.s. of \eq(\ref{eom2})
are
\be
\left(J_{\rm gf}\right)_a^\nu =
- \frac{1}{\xi} \partial^\nu \partial_\mu A^\mu_a
\ , \ \ {\rm and}\ \ \
\left(J_{\rm FP}\right)_a^\nu =
- g_s f_{abc} (\partial^\nu \bar{\et}_b ) \et _c \ . \label{Jgf} \ee

A generic operator vanishing by the classical EOM \eq(\ref{eom2})
has the form
\be
\OEOM \equiv \Y\cdot D_\m G^\mn
-O_{\qb q}-O_{\rm FP}-O_{\rm gf}\ ,\label{ODG}\ee
where $\Y$ denotes an arbitrary combination of further fields at the
same point and $O_{\qb q} \equiv \Y\cdot J_{\qb q}$, etc.
$D_\m G^\mn$ consists of a one-, a two-, and a three-gluon
piece. The one-gluon piece, $\partial_\m\partial^\m A^\n -
\partial^\n\partial_\m A^\m$, can be written in the
form $O_0 + J_{\rm gf}$. Then, $J_{\rm gf}$ cancels against the last
term on the r.h.s. of \eq(\ref{ODG}),
and the remainder $O_0$ is proportional to the inverse gluon propagator
($q^2 g_\mn - \frac{\xi-1}{\xi} q_\m q_\n$ in momentum space).

To illustrate \eq(\ref{eom0}) for $\OEOM$ of eq.~(\ref{ODG}), we
investigate first the (regularized) diagrams which have a vertex from
$O_0$. The external legs of the diagrams are given by the elementary fields
in \eq(\ref{eom0}) [with $\Phi\equiv A^\m_a$] which are on shell for
physical matrix elements. If the gluon field in $O_0$ is contracted
with an external gluon, the diagram contributes to the contact terms
on the r.h.s. of \eq(\ref{eom0}). The inverse propagator in $O_0$ acts
on the external leg and yields zero for on-shell matrix elements.
Possible diagrams in which the gluon field in $O_0$ is contracted with a
gluon in $\Y$ at the same point vanish in dimensional
regularization\footnote{
   The vanishing of these `tadpole'-like diagrams is a crucial
   property of dimensional regularization needed here; otherwise,
   additional vacuum subtractions are necessary.}.
In all other diagrams with a vertex from $O_0$ (working at first order
in $\Leff$!), the gluon propagator attached to $O_0$ ends in an
usual QCD-vertex, $V$, and is effectively canceled by the inverse
propagator in $O_0$.

These diagrams, with a vertex from $O_0$ next to a QCD-vertex $V$,
are in one-to-one correspondence to diagrams with one of the remaining
vertices from $\OEOM$: A diagram with $V$ being a three- or a
four-gluon vertex cancels with the corresponding diagram containing the
two- or the three-gluon piece of $D_\m G^\mn$, respectively (see fig.~1).
Finally, if $V$ is a quark-gluon or a ghost-gluon vertex, then the diagram
is canceled by the corresponding one with a vertex from $O_{\qb q}$
or $O_{\rm FP}$, respectively.

Similar considerations can be readily carried out for an operator which
vanishes by the classical EOM for the fermions (quarks),
\be\i\Ds\Psi \equiv
\i \left( \ds - \i g_s \As^a \frac{\l^a}{2} \right)\Psi = m\Psi \ ,
\label{eom1} \ee
or for the case of scalars (see also \cite{Collins}) and ghosts.

The main technical complication in proving the renormalized version of
\eq(\ref{eom}) or (\ref{eom0}) by means of diagrams arises from the fact
that bare graphs which are in one-to-one correspondence (in the sense
that they cancel each other) can have a different 1PI structure. Thus,
their `forests', which prescribe the renormalization, are not in
one-to-one correspondence, and it has to be shown that one can
remove recursively those forests which are not in one-to-one
correspondence \cite{EOM}.

\subsection{Terms from gauge-fixing and ghosts}
To study the relevance of the terms $J_{\rm gf} + J_{\rm FP}$
in the EOM for the gauge-boson field strength [see the r.h.s. of
eq~(\ref{ODG})], it is convenient to define two sets of operators:\\
$\OE$: Linear combinations of operators which vanish by
the classical EOM derived from (the basic part of) $\Lo$. Of course,
terms from gauge-fixing and ghosts in $\Lo$, and the EOM for the ghosts
themselves [see \eq(\ref{eom3})] are to be taken into account as well.\\
$\OB$: Linear combinations of operators which are the BRS variation of
some other operator.\\
The key observation for the discussion of the gauge-fixing and ghost terms
is the fact that $O_{{\rm gf}} + O_{{\rm FP}}$ belongs to
$\OB$ provided that $\Y^a_\n$ varies under BRS transformations as
\be \d \Y^a_\n = g_s f_{abc} \Y^b_\n \et^c \ ,
\label{dY}\ee
(with the shorthand notation $\d \equiv\d_{\rm BRS}/\d\omega$, where
$\d\omega$ is the infinitesimal parameter of the transformation). Then,
\be O_{\rm FP} + O_{\rm gf} = \d(- \Y^a_\m \partial^\m \bar{\et}_a)\ .
\label{dQ}\ee

Obviously, eq.~(\ref{dY}) holds  if $\Y^a_\n D_\m G^\mn_a$ is
gauge-invariant and contains no ghosts.
Typical examples for dimension six operators are $\Y^a_\n = D^\m G_\mn^a$ or
$\Y^a_\n = \bar{\Psi} \g_\n \l^a \Psi$,
where $\Psi$ may be left- or right-handed and a vector in flavour space.
After applying the EOM twice to $(D^\m G_\mn^a)(D_\l G^{\l\n}_a)$ one
arrives at $\Y^a_\n$ containing ghosts in the combination
$\Y^a_\n = (J_{{\rm gf}} + J_{{\rm FP}})^a_\n$. In this case an operator
from $\OE$, which vanishes by the EOM for the ghosts, has to be added on
the r.h.s. of \eqs(\ref{dY}) and (\ref{dQ}).

To see that an $\OBRS \in \OB$ does not contribute to physical (on shell)
matrix elements, one writes $\OBRS = \d\hat{O}$ and recalls (e.g.
\cite{Collins})
\bd \bra T \d\hat{O}(y) X(x_1,\ldots) \ket =
- \bra T \hat{O}(y) \d X(x_1,\ldots) \ket \ . \ed
The r.h.s. vanishes after LSZ reduction, because $\d X$ contains composite
fields which do not lead to a physical particle pole.

It is illustrating to compare the physical matrix elements of
$O_{\rm gf}$, for instance for $\Y^a_\n = \Psibar \g_\n T_a \Psi$,
with those of $J_{\rm gf}^{\n a}$.
The fact that matrix elements of $J_{\rm gf}^{\n a}$
{}~[$=\d\bigl(\partial^\n\bar{\eta}^a\bigr)\in\OB$] are zero is just the
familiar current conservation: The amplitude for
``$g^\ast \to$ physical on-shell particles'' vanishes when one
substitutes the polarization vector of the gluon $g^\ast$ by its momentum
(which can be on or off shell).
On the other hand, $O_{\rm gf}$ alone, which is not in $\OB$, has
non-vanishing matrix elements. This is readily understood for diagrams
in which the outgoing gluon line $g^\ast$ from $J_{\rm gf}$
closes a loop (e.g. in fig.~2a): At the $O{\rm gf}$-vertex the gluon
propagator is indeed contracted by its momentum; however, the diagram
corresponds to an amplitude where the $g^\ast$ `decays' into an off-shell
particle that again enters the $O_{\rm gf}$-vertex. The overall contribution
of these diagrams is canceled by the matrix element of $O_{\rm FP}$ in
fig.~2b (this can be verified directly by manipulation of the corresponding
diagrams \cite{diss}).

As to the renormalization of operators from $\OB$, one considers
an $\OBRS \equiv \d\hat{O}\in \OB$ and counterterms
$\hat{C}$, which render all Green's functions of $\hat{O}$ with
elementary fields finite. Then, the Green's functions
\bd \bra T\bigl(\OBRS+\d\hat{C}\bigr)(y)\cdot X(x_1,\ldots)\ket^\Ro =
- \bra T\bigl(\hat{O}+\hat{C}\bigr)(y)\cdot \d X(x_1,\ldots)\ket^\Ro\ ,\ed
where $\Ro$ indicates counterterms from $\Lo$,
can only be divergent if $y$ coincides with one of the $x_i$
(related with a composite field in $\d X$). Hence, only some operators
from $\OE$, together with $\d\hat{C}\in\OB$, are necessary as
counterterms for $\OBRS$.

\subsection{Reduction and RG evolution at first order in $\Leff$}
The stability of $\OE$ (respectively $\OE\oplus\OB$) under renormalization,
together with the fact that these operators have vanishing on-shell matrix
elements, implies that one may use the (naive) classical EOM to reduce
the basic part of an effective Lagrangian already before the RG evolution:
If the basic parts of two Lagrangians, renormalized at some scale $\mu$,
differ only by operators which are in $\OE$ ($\OE\oplus\OB$), then the
corresponding renormalized theories are on-shell equivalent,
and this property is preserved by the RG evolution to any other scale.

To demonstrate this in a more explicit way, we study the renormalization
of $\Leff$.
We assume that \eq(\ref{Leff1}) includes all linearly independent
operators which are allowed by global symmetries (Lorentz invariance,
flavour quantum numbers etc.) and by their (canonical) mass dimension.
Since Green's functions with {\it single} insertions of composite
operators require only counterterms with at most the dimension of the
inserted operator, $\Leff$ can be renormalized multiplicatively by
rewriting the effective Lagrangian in terms of renormalized
operators\footnote{
   Alternatively, the counterterms can be viewed
   as a renormalization of the couplings $c_i$; this approach is more
   convenient when working at higher order in $\Leff$ (see appendix~B).}
\be O_k^R = Z_{kl}\, O_l \ . \label{Z}\ee
Operator mixing arises when $Z$ is not a diagonal matrix, and we say that
an operator (or a set of operators) $O_k$ `mixes into' $O_l$ if
$Z_{kl} \neq 0$, i.e. if $O_k$ requires counterterms proportional to $O_l$.

Since $\OE$ mixes only into $\OE$, and since $\OB$ mixes only into
$\OB\oplus\OE$, $Z$ and consequently the anomalous dimension matrix
\be \g = Z \frac{d}{d\mu}\left(Z^{-1}\right) \ ,\label{gamma}\ee
have block-triangular form, when written in a suitable basis of
$\OE\oplus\OB$ and the remaining operators.
Therefore, after the RG evolution \footnote{The exponential is $g$-ordered.}
\be c_i(\mu_2) = \left[\exp \int_{g_s(\m_1)}^{g_s(\mu_2)}
           \frac{\g^T(g)}{\b(g)} dg \right]_{ij} c_j(\m_1) \ ,
\label{RG}\ee
the coefficients $c_i(\mu_2)$ of all $O_i\not\in\OE\oplus\OB$ do indeed
not depend on the initial values for the coefficients of the
operators from $\OE\oplus\OB$.

Finally, we note the useful result by Joglekar and Lee \cite{Jog} that
gauge invariant operators mix only among themselves and into $\OE\oplus\OB$.
Therefore, the operator basis for $\Leff$ can be restricted to
operators which are linearly independent of $\OE\oplus\OB$ {\it and}
gauge-invariant. Of course, `unphysical' operators which are not
gauge-invariant and which are not contained in $\OE\oplus\OB$ can mix
into physical operators. However, since the latter do not mix into
these unphysical operators, their coefficients remain zero if the
initial values vanish before the RG evolution.
This is also illustrated by a recent RG calculation of Grinstein and Cho
\cite{CG} where $O_{\rm FP}$ was included ``for completeness'':
$O_{\rm FP}$ is not in $\OE\oplus\OB$ and indeed mixes into gauge-invariant
operators.

\subsection{Equivalent effective Lagrangians at higher order}
When working beyond the linear approximation in the effective interactions
one obviously has to include the effective interactions themselves in
the adequate EOM (and one immediately wonders whether to use $\Lo + \Leff$
or $\Lo + \Leff- \OEOM$ in the derivation). However, simply removing
an operator which vanishes by the classical EOM, does {\it not} in general
lead to a `reduced'\footnote{
   Of course, one can always determine a `reduced' effective Lagrangian
   $\Lred$, which is on-shell equivalent to $\Leff$, by performing again
   the lengthy `matching' procedure with on-shell amplitudes to fix all
   coefficients for a reduced (e.g. ``canonical'' \cite{Arzt}) set of
   operators.}
effective Lagrangian which is on-shell equivalent to $\Leff$.
Green's functions with multiple insertions of $\OEOM$ (and $\Leff$)
have `non-trivial' contact terms which --- in contrast to those in
\eq(\ref{eom0}) --- do not vanish by LSZ reduction.

In practice, the effective couplings, collectively denoted by $\geff$,
are to be taken into account up to a certain order $M$ and the effective
Lagrangian has the form $\Leff
\equiv \geff\cdot\L_\eff^{(1)} + \ldots + {\geff}^M\cdot\L_\eff^{(M)}$.
If $\Leff$ contains an operator $O_\EOM^{(m)}$, which is of
order $\geff^m$, and which vanishes by the classical EOM derived from
$\Lo$, the `reduction' of the effective Lagrangian by the EOM amounts
essentially to the following task:
Starting from $O_\EOM^{(m)}$, construct a `reduced' effective Lagrangian,
\bd\Lred = \Leff - O_\EOM^{(m)} + O(\geff^{m+1})\ ,\ed
which is on-shell equivalent to $\Leff$ up to order $\geff^M$.

In the linear case, $M=1$, and for $m=M$ one simply has to remove
$O_\EOM^{(m)}$ from $\Leff$. Otherwise, the first step in constructing
$\Lred$ is to extend $O_\EOM^{(m)}$ recursively to an operator
$\OEOM^\prime$ which vanishes by the EOM derived form $\Lo + \L_\eff^\prime$.
Thereby, $\L_\eff^\prime$ is defined iteratively by rewriting $\Leff$ as
\be \Leff = \L_\eff^\prime + \OEOM^\prime \ . \ee
The iteration starts with
\bea
\L_\eff^\prime & = & \L_\eff - O_\EOM^{(m)} - O(\geff^{m+1})\ ,\nonumber \\
\OEOM^\prime & = & O_\EOM^{(m)} + O_\EOM^{(m+1)} + O(\geff^{m+2})\ ,
\nonumber \eea
such that $O_\EOM^{(m)}+O_\EOM^{(m+1)}$ vanishes by the EOM from
$\Lo+\L_\eff^\prime$ (keeping in $\L_\eff^\prime$ only terms of order
$\geff$).

We may assume
that $O_\EOM^{(m)}$ contains inverse propagators only through powers of
$\dEOM\Lo$, where $\dEOM\equiv \d/\d\Phi
- \partial_\mu \, \d/\d(\partial_\mu \Phi)$. Then, $\OEOM^\prime$ can be
chosen to contain inverse propagators only through powers of
$\dEOM(\Lo+\L_\eff^\prime)$. For instance, for
$O_\EOM^{(m)} = Q\cdot \left( \dEOM\Lo\right)^2$ one would set
$O_\EOM^{(m+1)} = 2 Q\cdot \dEOM\Lo \cdot \dEOM\L_\eff^{(1)}$,
etc.

Before removing $\OEOM^\prime$ we have to take into account its
on-shell contributions through non-trivial contact terms
on the r.h.s. of \eq(\ref{eom}) [or its iteration in the case that
$\OEOM^\prime$ contains higher powers of $\dEOM(\Lo+\L_\eff^\prime)$].
The amplitudes are evaluated from the first $M$ terms of the
Gell-Mann-Low series
\bea
\bra T \exp \Bigl\{ \i \int \Leff(y)\, \dy \Bigr\} X(x_1,\ldots)\ket
& = & \sum_{n=1}^M \frac{\i^n}{n!} \int dy_1\ldots dy_n
\bra T \Leff(y_1)\ldots\Leff(y_n) X(x_1,\ldots)\ket \nonumber \\[3mm]
& + & O(\geff^{M+1}) \ ;
\label{GML} \eea
and all on-shell contributions of $\OEOM^\prime$ remaining on the total
r.h.s. of \eq(\ref{GML}) correspond to \mbox{(sub-)}diagrams
with the following properties:
(a) All internal lines are canceled by inverse propagators in the
$\OEOM^\prime$ and there is at least one internal line,
(b) none of the inverse propagators in the $\OEOM^\prime$ acts on an
external line from $X(x_1,\ldots)$, and
(c) all vertices from $\Lo+\L_\eff^\prime$ have at least two lines,
which are canceled by inverse propagators from $\OEOM^\prime$.

Obviously, one can just choose suitable subdiagrams with property (a).
If (b) were not true, the diagram would not survive LSZ reduction;
and (c) is a consequence of (a) and the assumption that $\OEOM^\prime$
contains inverse propagators only through powers of
$\dEOM(\Lo+\L_\eff^\prime)$.

Sub-diagrams  with property (a) correspond to local vertices
which are obtained by shrinking the internal lines to a point. In this
way one can represent the contribution of each contact term by a local
operator, and  we denote the sum of all these `contact operators'
by $C[\OEOM^\prime]$.
More precisely, $C[\OEOM^\prime] = C_2 + C_3 + \cdots$, where the $C_n$
are determined by
\be\ba{l}
i \bra T C_n(y_1) X(x_1,\ldots) \ket
{\displaystyle\prod_{\nu=2}^n } \i \d(y_1-y_n)
\stackrel{!}{=}
\\ \hspace{2cm}
\left.\frac{\i^n}{n!} \bra T \OEOM^\prime(y_1) \ldots \OEOM^\prime(y_n)\,
\exp \Bigl\{ \i \int \L_\eff^\prime(y)\, \dy \Bigr\} X(x_1,\ldots) \ket
\right|_{(a)-(c)} \ \ \ ,
\ea \label{ntctct} \ee
for arbitrary elementary fields, $X(x_1,\ldots)$, but with the r.h.s.
restricted to the contributions of maximal connected diagrams
satisfying (a)--(c). Working up to order $\geff^M$, only a finite number
of contact operators appears in $C[\OEOM^\prime]$.

Finally, the `reduced' effective Lagrangian is obtained by replacing
$\OEOM^\prime$ by its contact operators
\be\Lred \equiv \L_\eff^\prime + C[\OEOM^\prime]\ .\ee
Note that $C[\OEOM^\prime]$ depends in a non-linear way on $\OEOM^\prime$
(or its coefficient, which we have absorbed in $\OEOM^\prime$).
If $\Leff = \L_\eff^\prime + \lambda\OEOM^\prime$, then
$\Lred = \L_\eff^\prime + C[\lambda\OEOM^\prime]$, which is given
in terms of the contact operators from \eq(\ref{ntctct}) by
$C[\lambda\OEOM^\prime] = \lambda^2 C_2 + \lambda^3 C_3 + \cdots$.

To illustrate the procedure, we consider a simple example:
$\Lo = $\mbox{$- \frac{1}{2} \Phi(\partial^2 + m^2)\Phi$}
$ + \Psibar\ds\Psi$ describes a massive scalar and a massless fermion,
and $\Leff = \geff \Psibar\Psi\partial^2\Phi$ are their (effective)
interactions. On-shell amplitudes at order $\geff$ (e.g. one scalar
decaying into two fermions) are obviously recovered by
$\L_\red^{(1)} = - \geff \Psibar\Psi m^2 \Phi =
\Leff - O_\EOM^{(1)}$, where $O_\EOM^{(1)} \equiv \geff \Psibar\Psi
(\partial^2+m^2)\Phi$.
At second order in $\geff$, $\Leff$ and $\L_\red^{(1)}$ are not
on-shell equivalent. In particular, $\Lo + \L_\red^{(1)}$ does not yield the
correct amplitude for the two-fermion scattering,
$A = - \i \geff^2 \frac{s^2}{s-m^2} + (s\leftrightarrow t)$, where $s$
and $t$ are the Mandelstam variables.
Following the above procedure, we set
\bd \OEOM^\prime = \O_\EOM^{(1)} + \frac{\d\Leff}{\d\Phi} =
\geff \Psibar\Psi (\partial^2+m^2)\Phi +
\geff^2 \Psibar\Psi \Psibar\Psi m^2 \ ,\ed
and from \eq(\ref{ntctct}), we find $C[\OEOM^\prime] =
\geff^2 /2\,\Psibar\Psi (\partial^2+m^2) (\Psibar\Psi)$.
One readily verifies that
$\Lred \equiv \Leff - \OEOM^\prime + C[\OEOM^\prime]$
indeed yields the same physical amplitudes as $\Leff$. In the
particular example at hand, this is the case even at arbitrary order in
$\geff$, because $\OEOM^\prime$ vanishes identically by the EOM derived
from $\Lo + \Leff - \OEOM^\prime$ and there are no further contact terms at
higher orders.

When working at the loop level, the above reduction procedure remains
essentially the same: All expressions are to be understood as
referring to the basic part of the renormalized (effective)
Lagrangian, and $C[\OEOM^\prime]$ is still given by the tree result
because loop diagrams with property (a) vanish in dimensional
regularization.  Since renormalized Green's functions of $\OEOM^\prime$
obey the same relations (derived from the basic part of the Lagrangian)
as the bare ones, $\L_\red^R$ is again on-shell equivalent to $\L_\eff^R$;
and the RG evolution preserves this equivalence. In contrast to the linear
approximation in $\geff$, it would be hard to demonstrate this in an
explicit way because the reduction by the EOM, and the RG evolution
(see, for instance, ref.~\cite{Shore}) are non-linear operations on
$\Leff$.

For the calculation of amplitudes at higher order in $\geff$, the Green's
functions with multiple insertions of $\Leff$ are actually needed only after
integration over the positions of the effective operators [see
\eq(\ref{GML})] and not in their general form. Therefore, it is sufficient
and in fact more convenient to view the counterterms for $\Leff$ as a
renormalization of the effective couplings (see appendix~B for more
details), rather than of the operators as in \eq(\ref{Z}).

\section{Effective treatment of rare B-decays}
We apply now the results of the previous section to rare B-decays;
in particular, we consider hadronic (and in less detail also
radiative) $b\to s$ transitions. (The analogous $b\to d$ modes follow
simply from replacing $V_{ts}$ by $V_{td}$.)  Since the momenta and
masses of the external particles are at most of the order of the B-mass
(and therefore much smaller than the mass scales governing the
propagation of virtual $t$-quarks or $W$-bosons), it is possible to
treat these processes in the framework of an effective low energy
theory \cite{EFT}.  The phenomenological procedure in this framework consists
of three steps: The derivation of the adequate effective Lagrangian,
the RG evolution and finally, the evaluation of the
hadronic matrix elements of the effective operators.

In step one, the heavy particles are integrated out. The resulting
effective action contains only the fields of the light particles,
however, possibly in a non-local way.  In order to obtain a local
effective Lagrangian one has to perform an operator product expansion
\cite{OPE} in the effective action.  If strong (and higher order
electroweak) interactions are absent, this simply corresponds to a
Taylor expansion in the external momenta.  Otherwise, the coefficient
functions of the local operators in the effective Lagrangian are
determined by a `matching' condition: Appropriate amplitudes
calculated in the effective theory are equated with those of the full
theory.  As soon as divergent loop diagrams of effective operators are
involved a renormalization scheme has to be specified.  The
coefficients will then depend on this scheme and, in particular, on
the renormalization scale $\m$. A natural choice for $\m$ to perform
the `matching' is of the order of the heavy masses, $M_W$ in our case.
A scale of this size allows to take into account (short-distance) QCD
effects perturbatively and minimizes logarithmic corrections from
higher order loops with heavy particles.

On the other hand, the matrix elements of the effective operators
involve logarithms of the ratio of $\m$ to the typical mass scale of
the low-energy process, which is of order $m_b$ in our case. These
large logarithms certainly distort a simple perturbative treatment of
the matrix elements and are removed by the RG evolution in step two:
The physical contents of the effective Lagrangian, `matched' at $\m
\approx M_W$, is translated to an effective theory with a
renormalization scale $\m \approx m_b$.  Thereby, the RG evolution of
the coefficients [see \eq(\ref{RG})] allows to improve the
(perturbative) treatment of short-distance QCD corrections by summing
up all powers of $\a_s \ell n \frac{\m^2}{M_W^2}$.

The RG evolution is straightforward in principle, however, the
explicit calculation is rather tedious and involves subtle
technical details.  There are numerous important contributions (see,
for instance, ref.~\cite{QCD} and [17--20]) and we refer to
Buras et al. \cite{Buras} and Misiak \cite{MM} for the most recent and
complete results. Here, we rather discuss some details of the derivation and
`matching' of the effective Lagrangian. This provides the starting
point for further phenomenological studies that intend to include the
RG improvement.

The third step, the evaluation of the matrix elements of the effective
operators with realistic hadron states is, of course, the most difficult
task because it requires genuine non-perturbative methods (or drastically
simplifying assumptions about the quark and gluon contents of the hadrons
together with some model to describe their binding effects). We shall
completely ignore non-perturbative effects and stay on the parton level
throughout.

\subsection{Effective vertices from penguin loops}
In this subsection, we describe one contribution to the effective
Lagrangian: Effective vertices resulting from one-particle-irreducible
(1PI) loop diagrams with internal top quark. The application of the EOM
already at this stage drastically simplifies further calculations
(e.g. of the `full' amplitudes required for the matching).
The `matching' procedure to properly define the complete effective
Lagrangian is postponed to the next subsection; only then, the effective
operators corresponding to $W$-exchange diagrams involving only light quarks,
like $b\to su\ub$ and $sc\cb$, will be introduced (and renormalized).

At leading order in $1/M_W^2$ the loop diagrams to be considered are
`penguins' with up to three gluons emitted from the $t$-line inside
the loop (diagrams with photons and leptons will be discussed below).
They are renormalized within the full theory by counterterms in the
original electroweak Lagrangian (which in fact cancel by the GIM
mechanism). As long as the penguins do not appear as subdiagrams
in two- (or more-) loop amplitudes, the external momenta are bound by
$m_b (\ll m_t$ or $M_W)$; and we can neglect any momentum dependence
which is higher order in $m_b^2/(m_t^2 {\ \rm or\ } M_W^2)$. Then, the
1PI diagrams are equivalent to (local) effective vertices and can be
summarized in terms of a `penguin' Lagrangian
\def\cp{c^\prime}
\def\ctp{\tilde{c}^\prime}
\def\Lp{{\cal {L}}_{\rm P}}
\be
\Lp = - 4 \frac{G_F}{\sqrt{2}} \, V_{tb}V^{\ast}_{ts} \,
\sum_{i} \cp_i P_i \ .\label{penguin} \ee

We choose the following set of gauge-invariant operators of dimension up
to six (see appendix~C for additional details)
\def\LR{{(L,R)}}
\be
\ba{ccl}
P_1 & = & \frac{g_s}{16 \pi^2}
\cdot \Psis \gamma_{\nu}L\frac{\lambda^{a}}{2} \Psib
\cdot \left( D_{\mu}G^{\mu \nu}\right) _{a} \ ,
\nonumber \\
P_2 & = & \frac{g_s}{16 \pi^2}
\cdot \Psis \sigma _{\mu \nu}\left( m_{b}R+m_{s}L\right)
\frac{\lambda ^{a}}{2}\Psib
\; G^{\mu \nu}_{a} \ \ \ (\equiv O_g)\ , \
\nonumber \\
P_3 & = & \frac{g_s}{16 \pi^2}
\cdot \Psis \Bigl\{ \i \not \! \! D \, , \,
\sigma_\mn G_a^{\mn} \frac{\lambda ^{a}}{2} \Bigr\} L \Psib \ ,
\nonumber \\
P_4 & = & \frac{g_s}{16 \pi^2}
\cdot \Psis \Bigl[ \i \not \! \! D \, , \,
\sigma_\mn G_a^{\mn} \frac{\lambda ^{a}}{2} \Bigr] L \Psib \ ,
\nonumber \\
P_5 & = & \frac{1}{16 \pi^2}
\cdot \Psis \i \not \! \!D  \not \! \!D \not \! \!D L \Psib \ ,
\nonumber \\
P_6 & = & \frac{1}{16 \pi^2}
\cdot \Psis \not \! \!D \not \! \!D
\left( m_{b}R+m_{s}L\right) \Psib \ ,
\nonumber \\
P_{7\LR} & = & \frac{1}{16 \pi^2}
\cdot \Psis \i \not \! \!D M_W^2 \LR \Psib \ ,
\nonumber \\
P_{8\LR} & = & \frac{1}{16 \pi^2}
\cdot \Psis M_W^2 (m_s L, m_b R) \Psib \ ,
\ea
\label{operatorbasis}
\ee
where $\Psib$ and $\Psis$ denote the quark fields, and the covariant
derivatives are defined in \eqs(\ref{eom2}) and (\ref{eom1}).

The coefficients $\cp_i$ are functions of the variable $x =
\frac{m_t^2}{M_W^2}$ (see appendix~C), and $\cp_{7L,R}$ and
$\cp_{8L,R}$ depend also on the renormalization scheme of the full theory.
Since the $b$- and $s$-momenta enter $P_4$ in an antisymmetric way,
$\cp_4$ is suppressed by an additional factor of the order $m_b^2/M_W^2$
and can be neglected.  We also note that only $\cp_1$ has a logarithmic
term, $\cp_1 \sim - \frac{2}{3} \ell n x + O(x)$, which would dominate
for $x\too 0$. All other coefficients approach a constant value in this
(unrealistic) limit.

All coefficient functions $\cp_i$ are uniquely determined by the (off-shell)
digrams for the $\sb b g$ vertex and the $\sb b$ self energy;
a calculation of the 1PI diagrams for $b\to sgg$ and $b\to sggg$ is not
necessary.  In turn, $\Lp$ provides the 1PI vertices for all $b\to s$
transitions involving up to three gluons (effective operators for more
than three gluons have dimension higher than six and are suppressed by
powers of $1/M_W^2$ or $1/m_t^2$).
This is simply a consequence of gauge invariance which is incorporated
in $\Lp$ by using manifestly gauge invariant operators. In momentum space
the corresponding relations among the various effective vertices are much
less transparent and the first complete treatments of $b\to sgg$ \cite{gg0}
exploited lengthy Slavnov-Taylor identities or a generalization of Low's
low energy theorem.

When $\Lp$ enters only at first order, we can take advantage of the EOM
(\ref{eom2}) and (\ref{eom1}), and considerably reduce $\Lp$: $P_5$,
$P_6$ and $P_{7L,R}$ become proportional to the flavour off-diagonal
mass terms $P_{8L,R}$ with coefficients that cancel when the full
theory is renormalized on shell\footnote{
   The on-shell renormalization conditions
   are equivalent to the requirement that all
   flavour off-diagonal mass terms vanish already when the EOM are
   applied either only to $\Psis$ or only to $\Psib$.}.
Applying the EOM to $P_1$, one obtains the four-quark operator
\bd O_P = \Psis\gamma_\nu L\frac{\lambda^a}{2}\Psib
\sum_{quarks}\qb\g^\nu\frac{\l^a}{2}q \ed
(plus additional operators from gauge-fixing and ghosts, which
may be dropped according to the discussion of section~2).
Finally, $P_3$ and $P_4$ yield color-magnetic dipole operators
equal to $P_2$
($P_4$ yields actually a different chiral structure
if $m_s$ is not neglected).

After this reduction with the help of the EOM, $P_2\equiv O_g$ is the
only operator which contains a gluon field; and $O_P$ and $O_g$ are the
only two operators remaining in $\Lp$. Their coefficients
\bea
\cp_P & = & - \frac{g_s^2}{16 \pi^2} \cp_1 =
- \frac{g_s^2}{16 \pi^2} \Bigl( F_1(x) + \frac{1}{9} \Bigr)\ , \nonumber \\
\cp_g & = & \cp_2 + \cp_3 = - \2 \Bigl( F_2(x) - \frac{2}{3} \Bigr)\ ,
\label{cg}\eea
can be expressed in terms of Inami-Lim functions \cite{IL} from the
$\sb b g$-vertex [$F_1$ and $F_2$, as defined in appendix~D, have no
constant terms for $x\to 0$; hence the explicit constants in (\ref{cg}), which
is irrelevant in all physical applications because of the GIM mechanism].

Radiative decays require additional penguin diagrams with one, two
or three photons coupling to the $W$ and to the unphysical Higgs
inside the loop. (By choosing a nonlinear gauge for the electroweak
sector on can avoid diagrams with a photon-$W$-Higgs coupling. Otherwise,
these give rise to additional operators \cite{gg0} which either
vanish on shell, or have coefficients that are independent of
the internal quark mass and, therefore, cancel by to the GIM mechanism.) The
resulting effective vertices correspond to four new operators, $\tilde{P}_1
\ldots \tilde{P}_4$, which are obtained from $P_1 \ldots P_4$ by the
replacements $G_\mn^a \to F_\mn$, $\frac{\l^a}{2} \to 1$ and
$g_s \to e$.  Among their coefficients, only $\ctp_1$ has a
logarithmic behavior, $\ctp_1 \too - \frac{4}{9} \ell n x$, while all
others become constant for $x\too 0$. By applying the EOM to
$\tilde{P}_1 \ldots \tilde{P}_4$ (and to $P_3 \ldots P_{7\LR}$, whose
covariant derivatives include, of course, also the photon field)
all local vertices for $b\to g\g$ or $b\to s \g\g$ are removed;
the only operator containing a photon field is $O_\g \equiv \tilde{P}_2$
and its coefficient is an other well-known Inami-Lim function
\be
\cp_\g = \ctp_2 + \ctp_3 =
 - \2 \bigl( \tilde{F}_2(x) - \frac{23}{18} \bigr)\ .
\label{cx}\ee

At order $\a \cdot G_F$, additional penguin diagrams with $Z$-bosons
(decaying into leptons), and $W$-box-diagrams arise. The resulting
effective four-Fermi vertices couple $\sb b$ to neutrinos, and to the
vector and axial current of the charged leptons. $\cp_\g$ and the
coefficients of these semi-leptonic operators depend on the gauge
that is chosen in the electroweak sector.  Of course, all (four-Fermi)
operators that remain after applying the EOM to $\tilde{P}_1$, have
gauge-independent coefficients (see also refs.~\cite{IL,PBE}).

$\Lp$ turns out to be useful even for the treatment of
diagrams with light internal quarks. Neglecting terms of order
$\frac{q^2}{M_W^2} \ll 1$, where $q^2$ collectively denotes the
external momenta, one can always split the amplitude for a penguin
diagram with an internal quark $i=u$, $c$, $t$ into two pieces:
\def\DA{\D{\rm A}}
\be A^{(i)} =  \sum_k \cp_k\bigl(\frac{m_i^2}{M_W^2}\bigr)\cdot
\bra f | P_k | b \ket^{\rm tree} + \DA\bigl(\frac{q^2}{m_i^2}\bigr)\ .
\label{splitting}\ee
For the heavy internal top quark, only the first term, arising from the
effective vertices in $\Lp$, is relevant. $\DA$ vanishes with powers of
$q^2/m_i^2$ for momenta small compared to the internal quark mass.
For light internal quarks the coefficients in the first (local) term
are evaluated at the appropriate value of $m_i^2$, while
the remaining momentum dependence of the amplitude is contained
in $\DA$ (including, for instance, threshold singularities in
$q^2/m_i^2$).

$\DA$ being the difference of the amplitudes at different
external momenta is worked out most conveniently by using the
four-Fermi approximation instead of the full $W$-propagator;
no UV divergence arises in $\DA$ and no renormalization is necessary.
For $b\to sq\qb$ \footnote{
   Eqs.~(\ref{splitting}) and (\ref{DeltaA}) should be contrasted to
   the more complicated result of ref.~\cite{Tanimoto}, which keeps
   terms of order $q^2/M_W^2$, but is not applicable for large
   $m_i \stackrel{>}{\sim} M_W$. In \eq(\ref{splitting}) the Inami-Lim
   function from $\cp_P$ gives always the correct mass dependence.}
\be \DA\bigl(\frac{q^2}{m_i^2}\bigr) = \D F_1\bigl(\frac{q^2}{m_i^2}\bigr)
\cdot \bra sq\qb|O_P|b\ket^{\rm tree} \ \label{DeltaA} \ee
differs from the local part just by a momentum dependent form factor
$\D F_1$ (see appendix~D).
In general, $\DA$ consists of matrix elements of local operators, some
of which may have dimensions higher than six, multiplied by momentum
dependent form factors.

The decomposition \eq(\ref{splitting}) is very convenient for
processes like $b\to sgg$ etc., where the momentum dependence $\DA$
contributes significantly in large regions of phase space and, hence,
can not be neglected. In the effective theory the momentum dependent
part $\DA$ is recovered by loop-level matrix
elements of effective four-Fermi operators.  However, due to their
renormalization, the correspondence is not straightforward and has to be
clarified in order to incorporate correction factors from
short-distance QCD in a systematic way.  To this end, we take a closer
look at the `matching' conditions that define the coefficients of the
operators in the effective Lagrangian.

\subsection{Matching of the effective Lagrangian}
In addition to the vertices from $\Lp$ we consider now the effective
operators for the usual $W$-exchange. Writing the effective Lagrangian as
\be
\Leff = 4 \frac{G_F}{\sqrt{2}} \left(
V_{cb}V^{\ast}_{cs}{\cal L}_c + V_{ub}V^{\ast}_{us}{\cal L}_u \right)
\ , \label{Leff}\ee
we need to discuss only ${\cal L}_c = \sum_{k} c_k O_k^R$ in the
following; ${\cal L}_u$ is simply obtained from ${\cal L}_c$ by
replacing the charm field (and mass) everywhere in $O_k$
(and $c_k$). In fact, since $V_{ub}V^{\ast}_{us} \ll V_{cb}V^{\ast}_{cs}$,
one may neglect ${\cal L}_u$ in most phenomenological applications
not concerned with CP violation. Note that all terms proportional to
$V_{tb}V^{\ast}_{ts}$ (from internal $t$-quarks) are distributed to
${\cal L}_c$ and ${\cal L}_u$ via the unitarity of the CKM matrix.

The local four-Fermi limit of the usual tree-level $W$-exchange,
$O_2 = \sb_\a \gamma^\mu L c_\a \cdot \cb_\b \gamma_\mu L b_\b$,
mixes with the operator
$O_1 = \sb_\a \gamma^\mu L c_\b \cdot \cb_\b \gamma_\mu L b_\a$
with reversed color structure. Both mix into further
four-quark operators, $O_3 \ldots O_6$, which carry various color
and chiral structures\footnote{
   We are here not interested in the explicit expressions for these
   operators; they can be found in e.g. \cite{Buras}.
   Note, however, that the Fierz ordering of the operators should be
   kept fixed throughout all calculations in $d \neq 4$ dimensions.},
and are related to the penguin operator $O_P = \2 \bigl[O_4 + O_6 -
\frac{1}{N} ( O_3 + O_5) \bigr]$. At the two-loop level the four-quark
operators mix into the color-magnetic moment operator $O_g$
\cite{GSW,MM}.

According to the discussion of section~2, all operators which vanish
by the EOM or can be written as BRS variations are irrelevant in the
present physical application (first order in $\Leff$!). In this sense,
$O_1 \ldots O_6$ and $O_g$ form a complete set of the dimension-five
or -six operators for hadronic $b\to s$ transitions. In addition,
$O_\g$, and the semi-leptonic four-Fermi operators, and electromagnetic
analogs of $O_3 \ldots O_6$ have to be included in order to describe
radiative and semi-leptonic decays, or to take into account mixing at
order $\a_{\rm QED}$.

To obtain a systematic treatment of QCD corrections, the (yet unknown)
coefficients at $\m = M_W$ are expanded in powers of $g_s^2$
\bd c_i(M_W) \equiv c_i^{(0)} + g_s^2 c_i^{(1)} + \cdots \ .\ed
Electromagnetic couplings are included by rewriting
$e = g_s \cdot \frac{e}{g_s}$ and treating $\frac{e}{g_s}$ as an
independent (running) parameter.

If the operators are defined with appropriate factors of $1/g_s^2$
\cite{MM} they do not mix at zeroth order and the
expansion of the anomalous dimension matrix starts at order $g_s^2$
\bd \g_{kl} \equiv  g_s^2 \cdot \g^{(1)}_{kl}
 + g_s^4 \cdot \g^{(2)}_{kl} + \cdots\ .\ed
The RG evolution \eq(\ref{RG}) can be written as
\bd c_i(\m) = \left(E^{(LL)}_{ij}(\m) +  g_s^2 \cdot E^{(NLL)}_{ij}(\m)
+ \cdots \right)  c_j(M_W) \ ,\ed
where the evolution matrices $E^{}$ sum up all orders of the product
$\a_s \cdot log(M_W^2/\m^2)$. Since the logarithm may be large, this
product is not expanded in $g_s$, but rather treated as an independent
parameter.

In leading-log (LL) approximation all $O(g_s^2)$ corrections are
neglected, and the $c_i^{(0)}(M_W)$ are determined by the matching condition
\be A^{\rm full}(b\to f) \stackrel{!}{=}
c_i^{(0)}(M_W) \cdot \bra f | O_i^R | b \ket^{(0)} \ + \ O(g_s^2)\ .
\label{matching0}\ee
were the amplitude in the full theory, $A^{\rm full}$, and the matrix
elements, $\bra f | O_i^R | b \ket^{(0)}$, have to be evaluated to order
$g_s^0$ (or to order $g_s$ if $A^{\rm full} = O(g_s)$, like for $b\to sg$).

Using appropriate final states, \eq(\ref{matching0}) yields
\be\ba{c}
c_2^{(0)} = - 1 \ ,  \\[0.2cm]
c_g^{(0)} = \cp_g\bigl(\frac{m_t^2}{M_W^2}\bigr)
          - \cp_g\bigl(\frac{m_c^2}{M_W^2}\bigr)
          = - \2 F_2 \bigl(\frac{m_t^2}{M_W^2}\bigr)
          + O\bigl(\frac{m_c^2}{M_W^2}\bigr)\ , \\[0.2cm]
c_\g^{(0)} = \cp_\g\bigl(\frac{m_t^2}{M_W^2}\bigr)
           - \cp_\g\bigl(\frac{m_c^2}{M_W^2}\bigr)
          = - \2 \tilde{F}_2 \bigl(\frac{m_t^2}{M_W^2}\bigr)
          + O\bigl(\frac{m_c^2}{M_W^2}\bigr)\ , \ea
\label{cLL}\ee
while all other coefficients are zero at $\m = M_W$.

Penguin contributions are order $\a_s$ in the full theory, and do in LL
approximation not enter the effective theory at $\m = M_W$. However,
when evolving to $\m < M_W$, $O_2$ mixes into $O_3 \ldots O_6$
($\g^{(1)}_{23} = -\frac{1}{2N}\frac{2}{3}\frac{1}{8\pi^2}$ etc.),
and if one expands their coefficients to leading order in $\a_s$,
they combine to $O_P$  with the coefficient
\be c_P(\m) = \frac{\a_s(\m)}{4\pi} \cdot \frac{2}{3}
\ell n \Bigl(\frac{\m^2}{M_W^2}\Bigr) \cdot c_2^{(0)}(M_W) \
+ \ \a_s\cdot O(1) + O(\a_s^2)\ .\label{running_cp} \ee

\def\rdep{\xi_{\rm R}}
For three-body decays like $b\to sgg$ etc., and for CP-violating
asymmetries in charged B-decays, one-loop matrix elements of the
effective operators are of basic interest because they give rise to
significant momentum dependent contributions \cite{gg0}
and to the crucial absorptive phase of the amplitudes \cite{BSS,GH,cp0}.
As an example, we consider the matrix
element
\be \bra sd\db | O_2^R | b \ket^{(1)} = - \frac{\a_s(\m)}{4\pi} \left[
\Delta F_1 \Bigl( \frac{q^2}{m_c^2}\Bigr)
- \frac{2}{3} \Bigl( \rdep + \ell n \frac{m_c^2}{\mu ^2} \Bigr) \right]
\cdot \bra sd\db | O_P | b \ket^{(0)} \ , \label{me} \ee
which corresponds to a penguin diagram with an internal $c$-quark in
the full theory. Here, $q^2$ is the invariant mass of the $d\db$-pair
and $\rdep = 1$ for naive dimensional regularization with
$\overline{ms}$ subtraction. The one-loop matrix element (\ref{me})
differs from the pure momentum dependence $\Delta F_1$ of the full
amplitude [see \eq(\ref{DeltaA})] by additional terms, which originate
from the renormalization.  Eqs.~(\ref{running_cp})
and (\ref{me}) demonstrate how the RG evolution from $\m = M_W$ to
$\m \approx m_b$ moves the large logarithms from the matrix elements
into the coefficient functions.

If the effective theory is matched only with $O(g_s)$ precision and
if the RG evolution is performed only in LL approximation, one can not
expect that higher order effects, like the $O(\a_s)$ matrix element in
(\ref{me}), exactly reproduce the full theory.  The appearance of
$\rdep$ clearly indicates some arbitrariness, because it depends on
the details of the renormalization scheme.
On the other hand, the logarithmic mass dependence and the proper
momentum dependence is recovered already at this stage: Combining
\eqs(\ref{running_cp}) and (\ref{me}) yields for $b\to sd\db$
\be A^{\rm eff} \sim \\ \frac{\a_s}{4\pi}
\left[ O(1) - \frac{2}{3} \ell n \frac{m_c^2}{M_W^2}
 + \Delta F_1 \Bigl( \frac{q^2}{m_c^2}\Bigr) \right]
\cdot \bra sd\db | O_P | b \ket^{(0)}
+ c_g^{(0)}\cdot \bra sd\db | O_g | b \ket \ , \label{Aeff} \ee
which is to be compared with the full theory [see \eq(\ref{splitting})]
\be A^{\rm full} \sim \\ \frac{\a_s}{4\pi}
\left[ F_1\Bigl(\frac{m_c^2}{M_W^2}\Bigr)
     - F_1\Bigl(\frac{m_t^2}{M_W^2}\Bigr)
 + \Delta F_1 \Bigl( \frac{q^2}{m_c^2}\Bigr) \right]
\cdot \bra sd\db | O_P | b \ket^{(0)}
+ c_g^{(0)}\cdot \bra sd\db | O_g | b \ket \ . \label{A_full} \ee
Recalling $F_1(x) = -\frac{2}{3} \ell n x + O(x)$, the two expressions
agree\footnote{
   $m_t^2/M_W^2$ and consequently $\ell n\; m_t^2/M_W^2$ are considered
   as $O(1)$; this is consistent with the approximation that the top
   and the $W$ are integrated out at the same scale.}
within ``logarithmic precision''; they differ only by non-logarithmic
terms of order $\a_s\cdot 1$ and, of course, exhibit the same momentum
dependence.

To go beyond this precision, one has to include the $O(g_s^2)$ corrections
in the initial values of the coefficients and the RG evolution has to be
performed in next-to-leading-log (NLL) approximation. To this end, the
$\g^{(2)}_{kl}$ are needed to evaluate the evolution matrix $E^{(NLL)}$, and
the $c_i^{(1)}(M_W)$ are obtained by solving
\be A^{\rm full}(b\to f) \stackrel{!}{=} c_i^{(0)} \cdot
\left( \bra f | O_i^R | b \ket^{(0)}
+ g_s^2 \cdot \bra f | O_i^R | b \ket^{(1)} \right)
\ + \ g_s^2 c_i^{(1)} \cdot \bra f | O_i^R | b \ket^{(0)} \ + \ O(g_s^4)\ ,
\label{matching1}\ee
where $A^{\rm full}$ and the matrix elements
$\bra f | O_i^R | b \ket \equiv \bra f | O_i^R | b \ket^{(0)} +
g_s^2 \cdot \bra f | O_i^R | b \ket^{(1)} + O(g_s^4)$ are needed
to order $g_s^2$ [$g_s^3$ for $A^{\rm full} = O(g_s)$].

For the penguin operators, $O_3 \ldots O_6$, which arise at $\m = M_W$ only
in the combination $O_P$, the NLL matching condition (\ref{matching1}),
combined with \eqs(\ref{cLL}) and (\ref{A_full}), yields
\be
c_P^{(1)}(M_W) = \frac{\a_s(M_W)}{4\pi} \left[
F_1(\frac{m_c^2}{M_W^2}) - F_1(\frac{m_t^2}{M_W^2})
+ \frac{2}{3} \left( \rdep + \ell n \frac{m_c^2}{M_W ^2} \right) \right]\ .
\label{cp_at_MW} \ee
Similarly, one determines the (scheme dependent) $O(\a_s)$ corrections
to the coefficients of $O_{1,2}$ (see ref.~\cite{Buras}).
Note that the logarithms in (\ref{me}) and (\ref{cp_at_MW}) combine with
$\Delta F_1 \Bigl( \frac{q^2}{m_{i}^2}\Bigr)$ and
$F_1\Bigl(\frac{m_{i}^2}{M_W^2}\Bigr)$, respectively ($i=u,c$).
Therefore, $\bra sq\qb | O_2^R | b \ket^{(1)}$ and $c_P^{(1)}(M_W)$ remain
finite in the limit of negligible internal quark masses ($m_{i}\to 0$).

Since $O_g$ and $O_\g$ do not mix into the four-quark operators
$O_1\ldots O_6$, one can consistently restrict a NLL analysis to this
subset \cite{Buras}. At $\m \neq M_W$,
the $O(\a_s\cdot 1)$ scheme dependence entering via the initial values of
the coefficients [due to the renormalization of the one-loop matrix
elements $\bra f | O_i^R | b \ket^{(1)}$ in \eq(\ref{matching1})]
is canceled by the corresponding scheme dependence in $E^{(NLL)}$
(see Buras et al. \cite{Buras} for a detailed analysis of this issue).

In decay modes, like $b\to sg, s\g, sgg, sg\g$, etc., where tree-level
and one-loop matrix elements contribute at the same (leading) order in
$g_s$, all divergencies in the one-loop matrix elements of the four-Fermi
operators cancel\footnote{
   In fact, this is readily seen with the help of the EOM:
   The counterterms for e.g. $\bra sgg | O_2 | b \ket^{(1)}$ are
   proportional to the operator $P_1$; the latter is equivalent to
   $O_P$ by the EOM and contributes just to the renormalization of
   matrix elements like \eq(\ref{me}).},
and therefore, no renormalization scheme dependence remains
even in LL approximation. However, a NLL analysis would be helpful to
reduce the $\mu$-dependence\footnote{
   By construction of the RG evolution, the $\mu$-dependence must cancel
   order by order in $\a_s$. However, the RG evolution of the
   coefficients includes all orders of $\a_s$ times log, while the
   matrix elements are evaluated only to order $g_s$ ($g_s^3$) in
   LL (NLL) approximation.}
of the physical results, which is the
more pronounced the larger the QCD corrections are relatively to the
uncorrected results at $\mu = M_W$ (e.g. in $b\to s\g$ \cite{Ali}). For the
operators $O_g$ and $O_\g$, the $O(g_s^3)$ matching \eq(\ref{matching1})
requires the finite parts of all corresponding two-loop graphs in the full
theory, and the $\g_{kl}^{(2)}$ for $k=1\ldots 6$ and $l=g,\g$ have to be
determined from tree-loop diagrams. So far, such a calculation has not yet
been attempted and the mixing of the four-quark operators into $O_g$
($O_\g$) is carried out only with LL times $g_s$ ($e$) precision \cite{MM}.

\section{Discussion}
In this paper we have studied the validity of the equations of motion
within the framework of an effective field theory described by a
Lagrangian $\Lo + \Leff$, where the interactions from $\Lo$ (e.g.
QCD or QED) are to be treated at arbitrary order and at the loop level.
We conclude that, in fact, $\Leff$ can be rewritten by freely using the
naive classical EOM (i.e. without gauge-fixing and ghost terms) derived
from $\Lo$, provided one considers only on-shell matrix
elements at first order in $\Leff$.
In particular, the effective Lagrangian may be reduced by the EOM before
(RG improved) short-distance QCD corrections are evaluated.

Although this result is not surprising, the reasoning involves
non-trivial ingredients, which are in principle well-known in a general
context \cite{Jog,Collins}. The essential steps have been carried out
explicitly here:
Besides renormalization effects for operators which vanish by the
classical EOM, we have investigated the relevance of the additional
terms in the correct EOM which arise from the gauge-fixing and ghosts.
The possibility of dropping these terms is due to the fact that they
appear in a combination which is the BRS variation of some other operator;
such operators do not mix into physical operators during the RG
evolution and their physical matrix elements vanish identically.

On the other hand, when the effective interactions from $\Leff$ are to
be used at higher order, operators which vanish by the classical EOM
{\it do} indeed lead to physical effects. We have described an
explicit procedure to derive a reduced effective Lagrangian which is
on-shell equivalent to the original one also beyond the linear order
in $\Leff$. In this procedure an operator $\OEOM$, which vanishes by
the classical EOM, is replaced by suitable (physical) operators which
account for all non-vanishing contact terms in Green's functions with
multiple insertions of $\OEOM$.

Since the renormalization group evolution preserves the relations due to
the EOM, we note that, it is not necessary for the definition of an
(`on-shell') effective theory to specify a ``canonical form'' \cite{Arzt}
of the set of operators: Although the matching with on-shell amplitudes
determines only a class of on-shell equivalent effective Lagrangians,
each of them will yield the same physical results after the RG evolution
to any other scale.

The ghost operator, $O_{\rm FP}$, has also been discussed by
Grigjanis et al. \cite{GODS}. However, there, $O_{{\rm FP}}$
did not arise through the EOM but was introduced by replacing the
`penguin' operator $P_1$ ($\equiv O_7$ in the notation of ref.~\cite{GODS})
through a combination of $P_1$ and $O_{\rm FP}$.  This ad hoc step
was justified by gauge invariance arguments \cite{GODS,Sutherland}:
Transversality is demanded for the sum of the one-loop diagrams that
renormalize $O_7$, and $O_{\rm FP}$ is added in order
to compensate for a diagram where the two gluons from $O_7$ close a loop
by emitting one (external) gluon via the three-gluon vertex of QCD. This
reasoning is not really compelling since only the combination $O_{\rm
gf} + O_{\rm FP}$ is BRS invariant (see appendix~A); in fact, the
non-transversal terms in ref.~\cite{Sutherland} can be compensated by
counterterms which contain no ghosts but rather vanish by the EOM for
the quarks.  Our discussion shows that $O_{\rm FP}$ arises in a natural
way after the use of EOM \eq(\ref{eom1}) --- however, always in
combination with $O_{\rm gf}$.

The operator $O_{\rm gf}$ was investigated by Eeg and Picek \cite{Eeg}.
They claim that the $\qs q_\m$ piece of the penguin subdiagram gives
rise to double-log terms in the mixing of $O_2$ into $O_\g$, whereas
such terms are not reproduced in the effective Lagrangian approach
\cite{GSW,GODS}. We recall that the $\qs q_\m$ piece corresponds to
the operator $O_{\rm gf}$ and note that there exist no two-loop diagrams
for $b\to s\g$ where $O_{\rm FP}$ contributes. Thus, our result
concerning the absence of physical effects from $O_{\rm gf} + O_{\rm FP}$
implies that the $\qs q_\m$ pieces, and the double-logs, in fact
cancel for the sum of all diagrams at hand, see fig.~4.  (Indeed, this
can be readily verified by standard Ward-Takahashi identities for the
gluon-quark vertex.)

Besides the striking advantage of the EOM for calculations of
short-distance QCD corrections, we find the EOM to be particularly
useful to keep track of the effective (penguin induced)
couplings that depend logarithmically on the internal quark masses.
(Recall that the weak GIM suppression of such terms is, for instance,
responsible for the amazing fact that a three-body decay like $b\to sq\qb$
has a larger branching ratio than the related two-body decay $b\to sg$.)
In the effective 'penguin' Lagrangian, $\Lp$, logarithmic coefficients
appear only for the operators $P_1$ and $\tilde{P_1}$, which contain the
gluon or photon field in the combination $D_\mu G^\mn_a$ or
$\partial_\m F^\mn$, respectively; via the EOM they are equivalent to
four-Fermi operators. Consequently, working at lowest order in $G_F$
and $\alpha_s$, the amplitudes for any $b\to s$ transition,
can receive $\ell n x$-terms ($x=m_i^2/M_W^2$) {\it only} through
contributions from the local four-Fermi operators
$\sb \g_\m L \frac{\lambda_a}{2} b \cdot \qb \g^\m \frac{\lambda_a}{2} q$ \
\ or\ \ $\sb \g_\m L b \cdot \bar{\ell} \g^\m \ell$. From this point of
view, the crucial --- but easily missed ---  cancellations of the
$\ell n x$-terms in $b\to sgg$ etc. \cite{gg0} become rather obvious.

The distinction between local and momentum dependent parts of the
amplitudes, which is convenient for purely calculational reasons in the
full theory, arises in a natural way in the effective theory: There, the
two parts originate from tree-level and one-loop matrix elements of
the effective Lagrangian, respectively, and receive in general
different corrections when short-distance QCD effects are included.
Some processes, like $b\to sq\qb$, have contributions from
$O(1)$ tree-level matrix elements (of local penguin operators)
{\it and} from $O(\a_s)$ one-loop matrix elements (of usual four-Fermi
operators).  A systematic and unambiguous treatment of both kinds of
matrix elements requires the use of the next-to-leading results for the
QCD-corrections. If only leading-log QCD corrections are included, the
{\it real} parts of the amplitudes can be recovered only up to
renormalization scheme dependent constants (which, of course, do not
affect the imaginary parts). Actually, taking the `leading-log'
approximation literally, one can (or should) drop the one-loop
matrix elements for $b\to sq\qb$ entirely, and consequently,
all CP asymmetries become zero \cite{Fleischer}.
In other decay modes, like $b\to s\g, sg, sg\g$, etc.,
no renormalization scheme dependence remains even in LL approximation,
but a NLL analysis would be desirable to reduce the $\mu$-dependence of
the results.

\renewcommand{\thesubsection} {}
\subsection{Acknowledgements}
I would like to thank D.~Wyler and M.~Misiak for their helpful collaboration
in various stages of this work, and A.~Ali, Ch.~Greub, M.~L\"uscher,
T.~Mannel, M.~Simonius and H.~Spiesberger for valuable
discussions and comments. During part of this work I enjoyed kind
hospitality at the University of Z\"urich and support by the Swiss
National Science Foundation.
%------------------------------------------------------------------------
\renewcommand{\thesection} {}
\renewcommand{\thesubsection} {Appendix \Alph{subsection}:}
\renewcommand{\theequation} {A.\arabic{equation}}
\setcounter{equation}{0}
\newpage
\section{Appendices}
\subsection{Gauge-fixing and ghosts}
In the background field formalism, the gauge field is decomposed into a
background field $B_\m^a$ and the `quantum field' $A_\m^a$. The Lagrangian
derives from the usual Yang-Mills and matter Lagrangian by replacing
the gauge-field by $B_\m^a + A_\m^a$. To fix the gauge, one adds
\be {\cal L}_{\rm gf} = - \frac{1}{2\xi} (D_\m^{ab}[B]\: A^\m_b)^2 \ , \ee
where we define, for general $V_\m^a$ and $\Phi^a$,
\bd D_\m^{ab}[V]\: \Phi^b \equiv \
\partial_\m \Phi^a + g f_{abc} V_\m^b \Phi^c \ . \ed
The resulting ghost Lagrangian is
\be {\cal L}_{\rm FP} =
\bar{\eta}_a D_\m^{ab}[B]\:D^\m_{bc}[B+A]\:\eta_c \ , \ee
and the corresponding currents on the r.h.s. of \eq(\ref{eom2}) become
\bd
\left(J_{\rm gf}\right)_a^\nu =
- \frac{1}{\xi} D^\nu_{ab}[B]\: D^\mu_{bc}[B]\: A_\mu^c
\ , \ \ {\rm and}\ \ \
\left(J_{\rm FP}\right)_a^\nu =
- g_s f_{abc} (D^\nu_{bd}[B]\: \bar{\et}_d ) \et_c \ . \ed
The theory is invariant under the following BRS transformations
\def\do{}
\bea
\d A_\m^a & = & D_\m^{ab}[B+A] \,\eta_b\do\ , \nonumber \\
\d B_\m^a & = & 0\ , \nonumber \\
\d \eta_a & = & \frac{g}{2} f_{abc} \eta_b \,\eta_c \do\ , \nonumber \\
\d \bar{\eta}_a & = & - \frac{1}{\xi} D_{ab}^\m[B]\, A_\m^b \do\ , \\
\d \Psi   & = & \i g \frac{\l^a}{2} \Psi \,\eta_a \do\ , \nonumber \\
\d \overline{\Psi} & = & - \i g \overline{\Psi} \frac{\l^a}{2} \,\eta_a \do
\ . \nonumber \eea
For any $\Y^a_\mu$, which transforms as
$\d \Y^a_\mu = g f_{abc} \Y^b_\mu\,\eta_c$,
the operator
$\Y^a_\mu \cdot \left(J_{\rm gf}+J_{\rm FP}\right)_a^\mu$
can be written as a BRS variation
\bd
\Y^\mu_a \cdot \left(J_{\rm gf} +J_{\rm FP}\right)^a_\mu =
\d\left( - \Y^\mu_a \cdot D^{ab}_\m[B]\, \bar{\eta}_b  \right) \ . \ed

Since the currents transform as
\bd \ba{lll}
\d\left(J_{\rm FP}\right)_a^\mu & = &
g f_{abc} \left(J_{\rm gf}\;+\;J_{\rm FP}\right)_b^\mu \,\eta_c\,\do\ ,\\
\d\left(J_{\rm gf}\right)_a^\mu & = &
-\frac{1}{\xi} D_{ab}^\mu[B]\, D_{bc}^\nu[B]\, D^{cd}_\nu[B+A]\, \eta_d\ ,\do
\ea \ed
the BRS variation of $\Y^a_\mu\cdot\left(J_{\rm gf}+J_{\rm FP}\right)_a^\mu$
(and of $J_{\rm gf}$) vanishes upon using the EOM for the ghosts
\be
D_{ab}^\nu[B]\, D^{bc}_\nu[B+A]\, \eta_c = 0 \ .\label{eom3}\ee
To obtain the corresponding expressions in a usual covariant gauge,
one simply sets $B_\m^a=0$ everywhere in the above formulae.

\subsection{Renormalization at higher order in $\Leff$}
\renewcommand{\theequation} {B.\arabic{equation}}
\setcounter{equation}{0}
The non-linear source renormalization (see ref.~\cite{Shore} for more
details) provides a conceptually simple method for the renormalization
of Green's functions with multiple insertions of composite operators.
The generating functional
\bd
W = \log\int{\cal D}\, \Phi\;\exp\; i\int(\Lo +J^B_0\Phi +J^B_i O_i)dx\ ,\ed
includes sources for the elementary fields and for all composite operators
(including the unity-operator). The bare sources,
$J^B_i \equiv S_i[J^R_1,\ldots]$, are general functions of the
renormalized sources, $J^R_i(x)$, and their derivatives. Renormalized
(respectively bare) Green's functions with insertions of the $O_i$ are
given by functional derivatives of $W$  with respect to the renormalized
(bare) sources.

The bare sources have an expansion
\be S_i[J^R_1,\ldots] = \left(J^R_l +
L^{jkl}_{\bf{m}\bf{n}} \partial^{\bf{m}}J^R_j
\partial^{\bf{n}}J^R_k + O\Bigl( (J^R)^3 \Bigr) \right)
\left[\mu^{\e {\bf D}} \hat{Z}\right]_{li}\ , \label{S}\ee
where the $L^{jkl}_{\bf{m}\bf{n}}$ are power series of poles in $\epsilon$
with residues that depend on the couplings of $\Lo$ only; and
$\partial^{\bf{m}}$ is a condensed notation for possible coordinate
derivatives, e.g. $\partial^0=1$, $\partial^{\mu_1 \mu_2}=\partial^2/
\partial x_{\mu_1}\partial x_{\mu_2}$, etc.
The diagonal matrix $\mu^{\e {\bf D}}$ contains powers of $\mu^\e$ which
are needed to leave the mass dimensions of the $J^R_i$
$\epsilon$-independent and to keep $\hat{Z}$ dimensionless.
The matrix of renormalization constants, $Z\equiv\mu^{\e {\bf D}} \hat{Z}$,
is the same as in \eq(\ref{Z}) and provides only the renormalization of
Green's functions with a single insertion of an $O_i$.

To evaluate amplitudes at higher order in $\Leff\equiv\sum g^B_i O_i$,
only Green's functions integrated over the coordinates of the insertions
of $\Leff$ are needed. Therefore, the $ g^B_i$ can be viewed as additional
couplings in the action and their (bare) values follow from \eq(\ref{S}):
\bd g_i^B \equiv G_i(g^R_1,\ldots,\mu) = S_i[g^R_1,\ldots] \ . \ed
Since the $g^R_i$ are coordinate independent, only the non-derivative
terms (${\bf m}={\bf n}=0$) from eq.~(\ref{S}) contribute in $G_i$.
The $\mu$-dependence of the $g^R_i$ is governed by the beta functions
\bd \b_i \equiv \left.\mu\frac{d g^R_i}{d\mu}
\right|_{g_j^B = const.} =
- \left[ \frac{\partial G}{\partial g^R} \right]^{-1}_{ij}
\mu\frac{\partial G_j}{\partial\mu} \ .\ed
On the other hand, the RG equation for the (generating functional of) the
Green's functions
\bd \left(\mu \frac{\partial}{\partial \mu}
       + \b \frac{\partial}{\partial g_o^R}
       + \g J_0^R \frac{\d}{\d J_0^R}
       + \g_i[J^R_1,\ldots] \frac{\d}{\d J_i^R} \right) W = 0\ , \ed
requires also the knowledge of the derivative terms of eq.~\eq(\ref{S})
for the `anomalous dimension functions'
\bd \g_i[J^R_1(x),\ldots] \equiv
\mu \frac{d}{d\mu} S_i[J^R_1(x),\ldots] \ . \ed

Of course, at first order in the $g^R_i$, the beta functions of the
effective couplings are related to the anomalous dimension matrix
\eq(\ref{gamma}),
\bd \b_i = - \g_{ij} g^R_j + O\Bigl( (g^R)^2 \Bigr) \ , \ed
and the $\mu$-dependence of the ``coefficients functions'', $c_i(\mu)$,
of eq.~(\ref{RG}) is equivalent to the running of the effective couplings,
$\bar{g}_i = c_i(\mu) + O\Bigl((g^R)^2\Bigr)$, governed by the RG equation
\bd \left(\mu \frac{\partial}{\partial \mu}
       + \b \frac{\partial}{\partial g_o^R}
       + \b_i \frac{\partial}{\partial g^R_i} \right)
\bar{g}_j(\mu,g^R_o,g^R_1,\ldots) = 0\ . \ed

\subsection{Penguin operators}
\renewcommand{\theequation} {C.\arabic{equation}}
\setcounter{equation}{0}
The operators $P_1\ldots P_{8L}$, as defined in \eq(\ref{operatorbasis}),
are linearly independent when ignoring relations due to the EOM. In fact,
together with the analogous operators having opposite chirality and with
the various four-quark operators, they form a complete set of gauge
invariant dimension five and six operators for hadronic $b\to s$ transitions.

The linear independence is readily verified from the Feynman-rules, which
we shall list here for the $b(k)\to s(k)$ self-energy, written as
$\frac{\i}{4 \pi^2} \frac{G_F}{\sqrt{2}} \S(k)$, and for the 1PI
vertex for $b(p)\to s(p') + g(\epsilon_a^\m,q)$, which we write as
$\frac{\i g_s}{4 \pi^2} \frac{G_F}{\sqrt{2}}\G_\m(p,p',q) \frac{\l^a}{2}
\epsilon_a^\m$. The contributions to $\S$ and $\G_\m$ from the various
operators are:\pagebreak[1]
\bd
\begin{tabular}{l|c|c}
$P_i$ & $\S(k)$ & $\G_\m(p,p',q)$ \\ \hline
$P_1$ & $0$ & $
\left( - q^2 \g_\m  + \qs q_\m \right) L$ \\
$P_2$ & $0$ & $
- 2\i \s_\mn q^\n \left(m_b R+m_s L\right)$ \\
$P_3$ & $0$ & $
2 \left[ \left( p^2 + p'^2 \right) \g_\m
-\ps p_\m -\ps p'_\m -\ps' p_\m - \ps' p'_\m + 2 \ps'\g_\m\ps \right]L$ \\
$P_4$ & $0$ & $
2 \left[ \left( - p^2 + p'^2 \right) \g_\m
+ \ps p_\m + \ps p'_\m - \ps' p_\m - \ps' p'_\m \right] L$ \\
$P_5$ & $ - k^2 \ks L$ & $
\left( - p^2\g_\m - p'^2\g_\m -  \ps'\g_\m\ps \right) L$ \\
$P_6$ & $- k^2 \left(m_b R+m_s L\right)$ & $
\left( - \g_\m\ps - \ps'\g_\m \right) \left(m_b R+m_s L\right)$ \\
$P_{7R}$ & $M_W^2 \ks R $ & $ M_W^2 \g_\m R$ \\
$P_{8R}$ & $M_W^2 m_b R$ & $0$ \\ \hline
\end{tabular}
\ed
\\
The form factors, which enter in a general decomposition of $\S$ and $\G_\m$,
\bea
\Sigma(k) & = & \left( c_{0}M^{2}_{W} + c_{1}k^{2}\right) m_{s}L + \left(
d_{0}M^{2}_{W}+d_{1}k^{2}\right) m_{b}R \nonumber \\
 & & + \left( e_{0}M^{2}_{W} + e_{1}k^{2}\right) \not \!kL + \left(
f_{0}M^{2}_{W}+f_{1}k^{2}\right) \not \!kR \ , \label{Sigma} \\
\Gamma_\mu(p,p',q) & = & \left[ \left( g_{00}M^{2}_{W} + g_{01}q^{2} +
g_{02}p^{2} + g_{03}p'^{2}\right) \gamma _{\mu} \right. \nonumber \\
 & & + \ \left. g_{1} \not \!p p_{\mu} + g_{2} \not \!p p'_{\mu} + g_{3}\not
\!p^{\,\prime}p_{\mu} \ + \ g_{4} \not \!p^{\,\prime}p'_{\mu} + g_{5} \not
\!p^{\,\prime}\gamma _{\mu}\not \!p \right] \cdot L \nonumber \\
 & & + \ \left( g_{6} \gamma _{\mu}\not \!p + g_{7} \not \!p^{\,\prime}\gamma
_{\mu} + g_{8} p_{\mu} + g_{9}p'_{\mu}\right) \cdot \left(
m_{b}R+m_{s}L\right) \label{Gamma} \nonumber \\
 & & + \ h_{0}M^{2}_{W} \gamma _{\mu}R \ ,\eea
can be found in ref.~\cite{gg0}. They satisfy relations
due to Ward-Takahashi identities,
\be\ba{lll}
g_{00} = e_0 \ , & \ \
h_{0} = f_0 \ , & \nonumber\\
c_1 = d_1 = g_7 + g_9\ , & \ \
e_1 = g_{01} + g_{02} + g_1\ , & \ \
f_1 = 0\ ,\nonumber\\
g_{01} = \frac{1}{2} \left( g_2 - g_1\right)\ , & \ \
g_{02} = \frac{1}{2} \left( g_2 + g_1\right) + g_{5}\ , & \nonumber \ea\ee
and due to the symmetry in $p$ and $p'$ (up to order $m_{b,s}^2/M_W^2$) ,
\bd g_{03} = g_{02} \ , \ \ g_3 = g_2 \ , \ \ g_4 = g_1 \ , \ \
g_7 = g_6 \ , \ \ g_9 = g_8 \ .\ed
The remaining form factors are in one-to-one correspondence with the
coefficient functions $\cp_i$ of the operators $P_i$
\be\ba{llll}
\cp_1 = \frac{1}{2} (g_2 - g_1) \ , & \ \
\cp_2 = - \frac{1}{2} g_8 \ , & \ \
\cp_3 = - \frac{1}{4} (g_1 + g_2) \ , & \nonumber \\
\cp_4 = 0 \ , & \ \
\cp_5 = g_1 + g_2 + g_5 \ , & \ \
\cp_6 = g_6 + g_8 \ , & \label{cprime} \\
\cp_{7L} = - e_0 \ , & \ \
\cp_{7R} = - f_0 \ , & \ \
\cp_{8L} = - c_0 \ , &
\cp_{8R} = - d_0 \ . \nonumber \ea\ee
The relations for photonic operators are similar: The 1PI
vertex for $b(p)\to s(p') + \g(\epsilon_\m,q)$, written in the form
$\frac{\i e}{4 \pi^2} \frac{G_F}{\sqrt{2}}\tilde{\G}_\m(p,p',q)\epsilon^\m$,
vanishes for $P_1\ldots P_4$, while $\tilde{\G}_\m = \G_\m$
for $\tilde{P}_1\ldots \tilde{P}_4$ and
$\tilde{\G}_\m = - \frac{1}{3} \G_\m$ for $P_5\ldots P_8$.
The $\ctp_1\ldots\ctp_4$ are obtained by replacing $\cp_i$ and $g_j$ in
\eqs(\ref{cprime}) by $\ctp_i$ and $\tilde{g}_j$, where the latter are
also given in \cite{gg0}.

\subsection{Inami-Lim functions}
\renewcommand{\theequation} {D.\arabic{equation}}
\setcounter{equation}{0}
For on-shell quarks (but off-shell gluon) the $\sb b g$-vertex
has only two independent form factors, the Inami-Lim functions $F_1$
and $F_2$ \cite{IL}:
\be
\Gamma_\mu(p,p',q) \ = \ F_1(x)(q^2\gamma_\mu-q_\mu \not\!q )L -
F_2(x) \i\sigma_{\mu\nu}q^\nu(m_b R + m_s L) \ .\label{Gamma_IL} \ee
After applying the EOM to the effective vertices in $\Lp$, all coefficients
can be expressed in terms of $F_1$ and $F_2$; their explicit expressions and
the limits for $x\too 0$ and $x\too 1$ are as follows
($\eta \equiv \frac{1}{x-1}$):
\def\limnull{\hspace{1.5cm} (x\too 0)\nonumber\\ }
\def\limone{\hspace{1cm} (x\too 1)\nonumber\\ }
\bea
F_1(x)
& = &   \frac{1}{12} - \2 \eta^3 + \frac{7}{12} \eta^2 + \frac{7}{6} \eta
      + \left( \2 \eta^4 - \3 \eta^3 - \frac{3}{2} \eta^2 \right) \ln x \\
& \too & - \frac{2}{3} \ln x + \frac{16}{3} x \ln x + \frac{3}{2} x
\hspace{-0.4cm} \limnull
& \too & \frac{43}{72} - \frac{19}{60} (x-1) \hspace{1.5cm} \limone
\nonumber \\
F_2(x)
& = & - \frac{3}{2} \eta^3 - \frac{9}{4} \eta^2 - \2 \eta + \frac{1}{4}
+ \left( \frac{3}{2} \eta^4 + 3 \eta^3 + \frac{3}{2} \eta^2 \right) \ln x \\
& \too & \2 x  \hspace{1.6cm} \limnull
& \too & \frac{1}{8} + \frac{1}{20} (x-1) \limone
\nonumber
\eea
For the $\sb b\g$-vertex $F_1, F_2$ have to be replaced by
$\tilde{F}_1, \tilde{F}_2$ in \eq(\ref{Gamma_IL}). The magnetic
moment, $\tilde{F}_2$, is given by
\bea
\tilde{F}_2(x)
& = &  \2 \et^3 + \frac{9}{4} \et^2 + \frac{29}{12} \et + \frac{2}{3}
      + \left(  - \2 \et^4 - \frac{5}{2} \et^3 - \frac{7}{2} \et^2
      - \frac{3}{2} \et \right) \ln x  \\
& \too & \frac{7}{12} x \hspace{1.75cm} \limnull
& \too & \frac{5}{24} + \frac{13}{120} (x-1) \limone \nonumber \eea
$\tilde{F}_1$ depends on the gauge in the electroweak sector, but combines
--- via the EOM --- with the coefficients of the semi-leptonic four-Fermi
operators to a gauge-invariant expression \cite{IL,PBE}.

The form factor of the momentum dependent part of the amplitude for
$b\to sq\qb$ [see \eqs(\ref{DeltaA}), (\ref{me})] is given by
\bea
\D F_1 (z) & = & -4 \int^1_0 u(1-u) \ell n \left[1-z u(1-u)\right] du
\nonumber \\[2mm]
& = & \frac{2}{3} \left\{ \frac{5}{3} + \frac{4}{z} +
      \bigl(1+\frac{2}{z}\bigr) R(z) \right\} \ ,\eea
where, setting $r\equiv \sqrt{|1-4/z|}$,
\bd
R(z) = \left\{ \ba{ll}
r\; \ln\frac{r-1}{r+1}               & (z<0) \\[1mm]
-2 + z/6 + z^2/60 + z^3/420 \cdots   & (z\to 0) \\[1mm]
2r\;{\rm arctan} (r) - r\pi          & (0<z<4) \\[1mm]
r\; \ln\frac{1-r}{1+r} +  \i r\, \pi & (z>4) \ea \right. \ed
%-------------------------------------------------------------------------

%-------------------------------------------------------------------------
\vspace*{2cm}
\section{Figure Captions}

{\bf Fig.~1:} Subdiagrams with insertions of $\OEOM$ of \eq(\ref{ODG}).
The crossed blob $\otimes$ denotes $O_0$, and the hatched ones represent
the two- and three-gluon pieces of $D_\mu G^{\mu\nu}$. The dots indicate
further fields in $\Y$.\\[1cm]
{\bf Fig.~2:} Diagrams which, in principle, may generate one-loop mixing of
$O_{\rm gf} + O_{\rm FP}$ into four-quark operators, or to finite
contributions to the matrix element for $b\to sq\qb$.\\[1cm]
{\bf Fig.~3:} Diagrams for the mixing of $O_{{\rm gf}} \sim \qs q_\mu$
into $O_\g$. The cross $\times$ indicates the various places where the
gluon from $O_{{\rm gf}}$ has to be attached to.
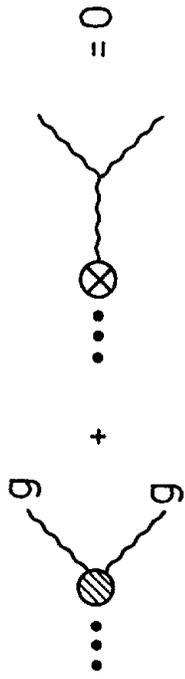
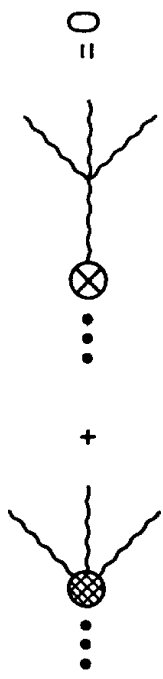
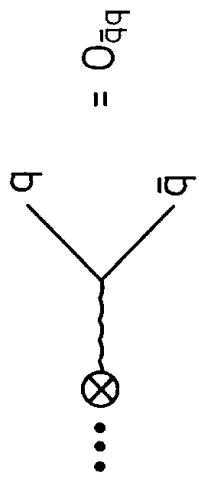
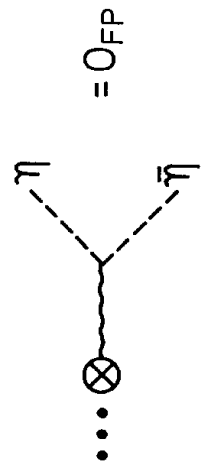
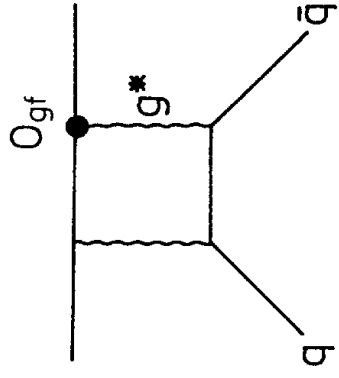
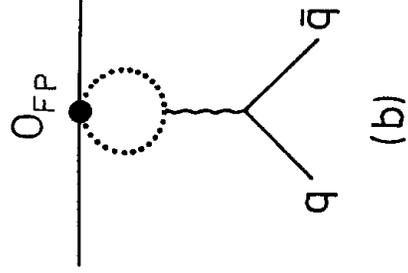
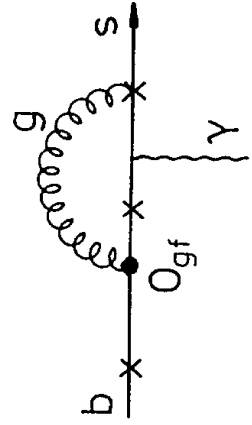
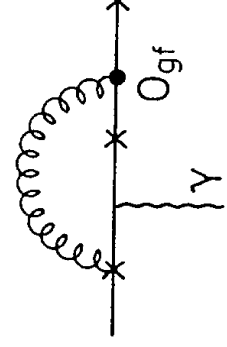
\end{document}